\documentclass[aps,prd,superscriptaddress,notitlepage,showpacs,showkeys,10pt]{revtex4-1}

\usepackage{graphicx}
\usepackage{amsmath}
\usepackage{amssymb}
\usepackage{color}
\usepackage{bm}

\newcommand{\sign}[1]{\,\mbox{sgn}\left({#1}\right)}

\definecolor{purple}{rgb}{0.8,0,0.6}
\definecolor{darkgreen}{rgb}{0.00,0.6,0.00}
\begin{document}
\title{Origin of Bardeen-Zumino current in lattice models of Weyl semimetals}

\author{E.~V.~Gorbar}
%\email{gorbar@bitp.kiev.ua}
\affiliation{Department of Physics, Taras Shevchenko National Kiev University, Kiev, 03680, Ukraine}
\affiliation{Bogolyubov Institute for Theoretical Physics, Kiev, 03680, Ukraine}

\author{V.~A.~Miransky}
%\email{vmiransk@uwo.ca}
\affiliation{Department of Applied Mathematics, Western University, London, Ontario, Canada N6A 5B7}

\author{I.~A.~Shovkovy}
%\email{igor.shovkovy@asu.edu}
\affiliation{College of Integrative Sciences and Arts, Arizona State University, Mesa, Arizona 85212, USA}
\affiliation{Department of Physics, Arizona State University, Tempe, Arizona 85287, USA}

\author{P.~O.~Sukhachov}
%\email{psukhach@uwo.ca}
\affiliation{Department of Applied Mathematics, Western University, London, Ontario, Canada N6A 5B7}

\begin{abstract}
For a generic lattice Hamiltonian of the electron states in Weyl semimetals, we calculate
the electric charge and current densities in the first order in background electromagnetic and strain-induced
pseudoelectromagnetic fields. We show that the resulting expressions for the densities contain
contributions of two types. The contributions of the first type coincide with those in the chiral kinetic theory.
The contributions of the second type contain the information about the whole Brillouin zone and cannot be
reproduced in the chiral kinetic theory. Remarkably, the latter coincide exactly with the Bardeen-Zumino
terms that are usually introduced in relativistic quantum field theory in order to define the consistent anomaly.
We demonstrate the topological origin of the Bardeen-Zumino (or, equivalently, Chern-Simons) corrections
by expressing them in terms of the winding number in the lattice Hamiltonian model.
\end{abstract}

%\pacs{03.65.Sq}
%71.45.-dCollective effects
%71.45.GmExchange, correlation, dielectric and magnetic response functions, plasmons
%03.65.SqSemiclassical theories and applications
%72.30.+qHigh-frequency effects; plasma effects

\maketitle

\section{Introduction}
\label{sec:introduction}

The possibility to observe the signatures of the chiral anomaly is one of the most intriguing aspects
of the physics of Weyl semimetals, whose low-energy quasiparticles are described by the corresponding
relativistic-like equation in the vicinity of Weyl nodes. A condensed-matter realization of Weyl fermions
was first predicted theoretically in pyrochlore iridates \cite{Savrasov}. Later, a number of materials
(e.g., $\mathrm{TaAs}$, $\mathrm{TaP}$, $\mathrm{NbAs}$, $\mathrm{NbP}$, $\mathrm{Mo_xW_{1-x}Te}$,
and $\mathrm{YbMnBi_2}$) were discovered to be Weyl semimetals
\cite{Weng-Fang:2015,Qian,Huang:2015eia,Bian,Huang:2015Nature,Zhang:2016,Cava,Belopolski}.
In accordance with the general arguments of Nielsen and Ninomiya \cite{Nielsen-Ninomiya}, Weyl
nodes in such condensed-matter materials come in pairs of opposite chirality. In fact, typical Weyl
semimetals have multiple pairs of opposite-chirality nodes in the reciprocal space
that are shifted from each other either in momentum or energy. The corresponding chiral
structure implies that the time-reversal symmetry or parity is broken. This is
also what makes them qualitatively different from the Dirac semimetals, such as A$_3$Bi ($\mathrm{A=Na,K,Rb}$), Cd$_3$As$_2$, and ZrTe$_5$ \cite{Weng,Wang,Weng:2014}, in which
pairs of opposite-chirality nodes overlap, and both discrete symmetries are preserved.

Formally, the low-energy effective theory of a Weyl semimetal is invariant under the chiral symmetry which is
generated by independent phase transformations of the left- and right-handed fermions. Such a symmetry,
however, is anomalous in the presence of parallel electric and magnetic fields \cite{ABJ}. In relativistic
quantum field theories, the origin of the anomaly is connected with the absence of an ultraviolet (high-energy)
regularization consistent with both electric and chiral charge conservation. In condensed matter systems,
on the other hand, the Brillouin zone is always finite in the reciprocal space and, thus, ultraviolet divergences are absent.
Because of its topological roots \cite{Nielsen-Ninomiya}, however, the chiral anomaly
is still present and can have observable implications. One of them is a large negative magnetoresistance
\cite{Aji:2012,SonSpivak,Gorbar:2013dha,Burkov:2015} that was experimentally observed in
$\mbox{Na}_3\mbox{Bi}$ \cite{Xiong}, $\mbox{Cd}_3\mbox{As}_2$ \cite{Li-Wang:2015,Li-He:2015},
$\mbox{Zr}\mbox{Te}_5$ \cite{Li},  $\mbox{Gd}\mbox{Pt}\mbox{Bi}$
\cite{Ong}, and $\mbox{TaAs}$ \cite{Zhang:2016}.

An efficient approach to study the electromagnetic response of Weyl semimetals is the
chiral kinetic theory \cite{Son:2012wh,Stephanov,Son:2012zy}, which is the generalization of the
standard kinetic theory \cite{Landau:t10} to the case of Dirac and Weyl quasiparticles. Not only
does it capture the topological properties of the chiral fermions via the Berry curvature \cite{Berry:1984} but also
correctly describes the chiral anomaly in background electromagnetic fields. It turns out, however,
that the chiral kinetic theory does not include all topological contributions relevant for Weyl semimetals.
In particular, it misses the Bardeen-Zumino current (also known as the Chern-Simons
current), which is critical for the correct description of the chiral magnetic effect \cite{Franz:2013,Basar:2014},
the anomalous Hall effect \cite{Ran,Burkov:2011ene,Grushin-AHE,Zyuzin,Goswami}, and collective
excitations in Weyl materials \cite{Gorbar:2016ygi}. Note that the Bardeen-Zumino term \cite{Bardeen}
was first proposed in relativistic quantum field theories in order to define the consistent anomaly
(for an instructive discussion of the Bardeen-Zumino current in the context of Weyl
semimetals, see Refs.~\cite{Landsteiner:2013sja,Landsteiner:2016}).

In this connection, let us briefly recall the concepts of covariant and consistent anomalies in the
high energy physics \cite{Landsteiner:2016}. Because of the regularization ambiguity in the
calculation of the triangle diagram \cite{ABJ}, there is a freedom in the definition of electric
and chiral current densities in quantum field theory with chiral fermions. The results are defined
up to a Bardeen--Zumino polynomial  \cite{Bardeen} in gauge fields. In the covariant scheme,
the currents are required to couple covariantly to background gauge fields. However, they cannot
be defined as functional variations of a quantum effective action. In the consistent scheme, the
corresponding currents are given by the variations of the quantum action. Such currents are
consistent with the local electric charge conservation even in the presence of both vector and axial gauge fields.

Theoretically, the absence of the Bardeen-Zumino current in the chiral kinetic theory becomes a
particularly acute problem in Weyl semimetals subjected to pseudoelectromagnetic fields induced
by mechanical strains
\cite{Zhou:2012ix,Zubkov:2015,Cortijo:2016yph,Cortijo:2016wnf,Grushin-Vishwanath:2016,Pikulin:2016,Liu-Pikulin:2016}.
 In essence, these fields resemble the
ordinary electromagnetic ones but couple to opposite chirality quasiparticles with different signs.
The standard formulation of the chiral kinetic theory would then imply a local nonconservation
of the electric charge in background electromagnetic and pseudoelectromagnetic fields. Clearly, this is
unacceptable. Therefore, the chiral kinetic theory should be amended by including the
additional Bardeen-Zumino terms in the definition of the charge and current densities \cite{Gorbar:2016ygi}.
In the four-vector notation, the explicit form of the corresponding fermion current reads $
j^{\nu}_{\text{{\tiny BZ}}} =  -e^2\epsilon^{\nu \rho \alpha \beta} A_{\rho}^5 F_{\alpha \beta}/(4\pi^2)$
\cite{Bardeen,Landsteiner:2013sja,Landsteiner:2016}, where field $A_{\rho}^5=b_{\rho}+\tilde{A}^5_{\rho}$
consists of the strain-induced pseudoelectromagnetic gauge field $\tilde{A}_{\rho}^5$ and the chiral shift
four-vector $b_{\rho}=(b_0,-\mathbf{b})$. Note that $b_0$ and $\mathbf{b}$ describe the energy and
momentum-space separation between the Weyl nodes, respectively.

In relativistic field theory, the structure of the Bardeen-Zumino term can be established
\cite{Bardeen,Landsteiner:2013sja} by requiring that the ultraviolet regularization
of the theory is consistent with the electric charge conservation. The same argument is often tacitly
followed in Weyl semimetals, although there are no ultraviolet divergencies in condensed matter
systems and the concept of chiral quasiparticles works only in the vicinity of the Weyl nodes.
%fails beyond the low-energy region of the theory.
Thus, one of the main goals of this study is to understand whether the Bardeen-Zumino current
is universal and topologically protected in realistic models of Weyl semimetals with a finite Brillouin
zone. As we will show, it is proportional to the winding number of the mapping of a two-dimensional section of the Brillouin zone onto the unit sphere and, thus, indeed has a topological
origin. Moreover, we will demonstrate that the result is quite general and works even in the case
of multi-Weyl semimetals \cite{Volovik:1988,Fang-Bernevig:2012,Li-Roy-Das-Sarma:2016}
(i.e., Weyl semimetals with the topological charge of the nodes greater than one),
whose topological responses were recently discussed in
Ref.~\cite{1705.04576} using the Fujikawa's regularization method.

This paper is organized as follows.
In Sec.~\ref{sec:lattice}, we introduce a generic lattice model of Weyl semimetals and outline the general
formalism that we will use to study the electromagnetic response. In Sec.~\ref{sec:lattice-first},
we derive the expressions for the electric charge and current densities in the linear order
in a background magnetic field and compare the results with their counterparts in the chiral
kinetic theory. We study the response to a background electric field in Sec.~\ref{sec:Kubo}.
The response to strain-induced pseudomagnetic and pseudoelectric fields is studied in
Secs.~\ref{sec:lattice-B5} and \ref{sec:Kubo-E5}, respectively. The summary of our results
and general conclusions are presented in Sec.~\ref{sec:Summary-Discussions}.
Technical details of derivations are given in several appendices at the end of the paper.
Throughout the paper, we use the units with $\hbar = c = 1$.

\section{Model}
\label{sec:lattice}

The electron states in a generic lattice Weyl semimetal can be described by the
following Hamiltonian \cite{Ran,Franz:2013} in the momentum space:
\begin{equation}
\label{lattice-d-def-be}
\mathcal{H}_{\rm latt} = d_0 +\mathbf{d}\cdot\bm{\sigma},
\end{equation}
where $\bm{\sigma}=(\sigma_x,\sigma_y,\sigma_z)$ are the Pauli matrices and
functions $d_0$ and $\mathbf{d}$ are periodic in quasimomenta $\mathbf{k}=\left(k_x,k_y,k_z\right)$.
The latter can have, for example, the following explicit form:
\begin{eqnarray}
\label{d-def-be}
d_0 &=& g_0 +g_1\cos{(a_zk_z)} +g_2\left[\cos{(a_xk_x)}+\cos{(a_yk_y)}\right],\\
d_1 &=& \Lambda \sin{(a_xk_x)},\\
d_2 &=& \Lambda \sin{(a_yk_y)},\\
d_3 &=& t_0 +t_1\cos{(a_zk_z)} +t_2\left[\cos{(a_xk_x)}+\cos{(a_yk_y)}\right],
\label{d-def-ee}
\end{eqnarray}
where $a_x$, $a_y$, and $a_z$ denote the lattice spacings and parameters $g_0$, $g_1$, $g_2$, $\Lambda$, $t_0$, $t_1$, and $t_2$ are material dependent. Their characteristic values valid for
$\mathrm{Na_3Bi}$ are given in Appendix~\ref{Sec:App-model} and will be used in our numerical calculations. Also, for the sake of simplicity, we will assume that the lattice is cubic, i.e., $a_x=a_y=a_z=a$.

\begin{figure}[!ht]
\begin{center}
\includegraphics[width=0.45\textwidth]{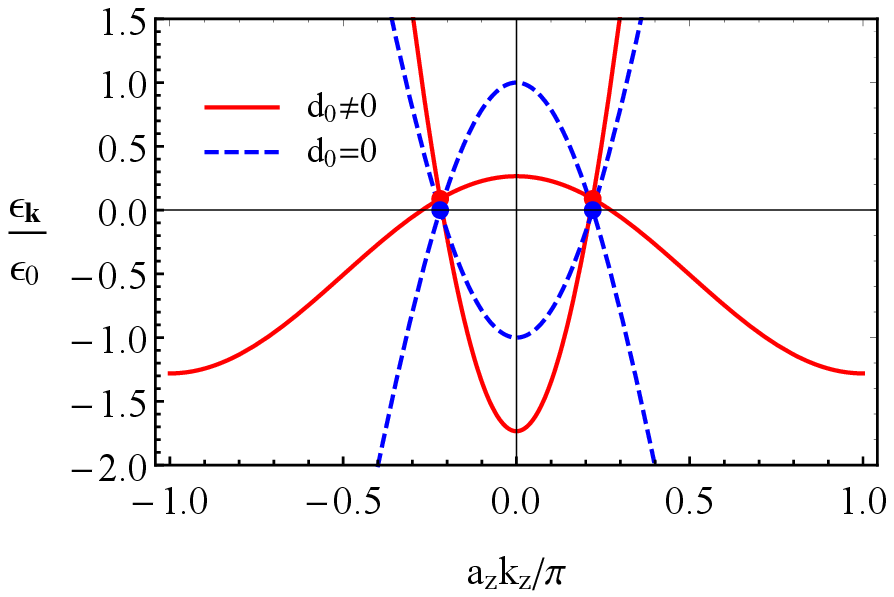}\hfill
\includegraphics[width=0.45\textwidth]{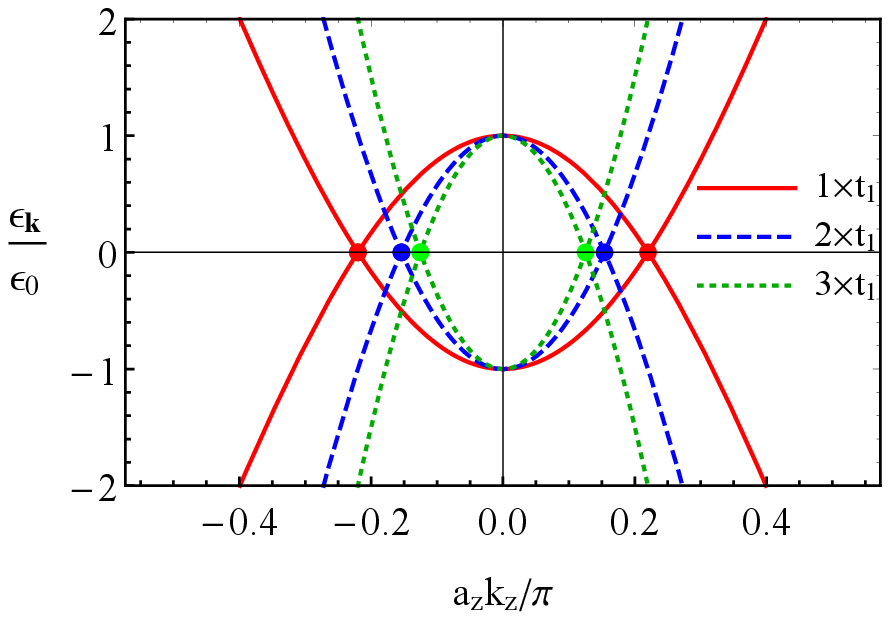}
\caption{Left panel: The energy spectrum of the lattice model (\ref{lattice-d-def-be})
given by Eq.~(\ref{lattice-E}) for $d_0\neq0$ (red solid lines) and $d_0=0$ (blue dashed
lines) as a function of $k_z$ at $k_x=k_y=0$. Right panel: The energy spectrum (\ref{lattice-E}) for
several different values of parameter $t_1$. The complete set of model
parameters is given in Appendix~\ref{Sec:App-model}.}
\label{fig:lattice-energy}
\end{center}
\end{figure}

As is easy to check, the dispersion relations of quasiparticles described by the model Hamiltonian (\ref{lattice-d-def-be})
are given by
\begin{equation}
\epsilon_{\mathbf{k}} =  d_0 \pm |\mathbf{d}|.
\label{lattice-E}
\end{equation}
When the parameters are such that $|t_0+2t_2|\leq |t_1|$, this model has two Weyl nodes separated
in momentum space by $\Delta k_z=2b_z$, where the chiral shift parameter $b_z$
is given by the following analytical expression:
\begin{equation}
b_z=\frac{1}{a_z} \arccos{\left(\frac{-t_0-2t_2}{t_1}\right)}.
\label{lattice-bz}
\end{equation}
For simplicity, we will assume that the energy vanishes at the position of Weyl nodes.
In terms of the model parameters, this implies that $g_0+2g_2-g_1(t_0+2t_2)/t_1=0$. In a general
case, this condition can be enforced by an appropriate redefinition of the reference point
for the chemical potential. Further, as one can see from the left panel of Fig.~\ref{fig:lattice-energy},
$d_0$ introduces an asymmetry between the valence and conduction bands, which complicates
the analysis. Therefore, for the sake of simplicity, in what follows we will drop the term $d_0$. It is
instructive to note, that $\epsilon_{0}$ [which is the value of energy (\ref{lattice-E}) at
$\mathbf{k=0}$] defines the height of the ``dome" in the energy spectrum. On the other hand,
parameter $t_1$ affects both the momentum space separation between the Weyl
nodes and the value of the Fermi velocity.

%%%%%%%%%%%%%%%%%%
%%%%%%%%%%%%%%%%%%

In order to study a linear electromagnetic response in the Weyl semimetal, we include an interaction with a
gauge field through the following interaction term:
\begin{equation}
\mathcal{H}_{\rm int} = \mathbf{j}\cdot\mathbf{A},
\label{lattice-H-int}
\end{equation}
where the electric current density operator in the momentum space is given by
\begin{equation}
\mathbf{j}(\mathbf{k}) = -e\bm{\nabla}_{\mathbf{k}} \mathcal{H}_{\rm latt},
\label{lattice-current-gen}
\end{equation}
and $e$ is a fermion charge. By using Eqs.~(\ref{d-def-be})--(\ref{d-def-ee}),
we derive the explicit expressions for the components of the current, i.e.,
\begin{eqnarray}
\label{lattice-coeff-j-be}
j_x&=& e\,a_x\left[t_2\sin{(k_xa_x)}\sigma_z +g_2\sin{(k_xa_x)} -\Lambda \cos{(k_xa_x)} \sigma_x\right],\\
\label{lattice-coeff-j-y}
j_y&=& e\,a_y\left[t_2\sin{(k_ya_y)}\sigma_z +g_2\sin{(k_ya_y)} -\Lambda \cos{(k_ya_y)} \sigma_y\right],\\
\label{lattice-coeff-j-ee}
j_z&=& e\,a_z\left[t_1\sin{(k_za_z)}\sigma_z +g_1\sin{(k_za_z)}\right].
\end{eqnarray}
In a many-body system, the electric charge and current densities are given in terms of the
quasiparticles Green's function $G(r,r^{\prime})$ as follows:
\begin{eqnarray}
\label{lattice-rho-def}
\rho &=& -e\lim_{r^{\prime}\to r}\mbox{tr}\left[G(r,r^{\prime})\right],\\
\label{lattice-J-def}
\mathbf{J} &=& -\lim_{r^{\prime}\to r}\mbox{tr}\left[\mathbf{j}(-i\bm{\nabla}_{\mathbf{r}})G(r,r^{\prime})\right],
\end{eqnarray}
where $r=(t,\mathbf{r})$ and $r^{\prime}=(t^{\prime},\mathbf{r}^{\prime})$. To the linear order in the background
electromagnetic fields, the Green's function has the form
\begin{equation}
\label{lattice-G-def-be}
G(r,r^{\prime}) = G^{(0)}(r-r^{\prime})+G^{(1)}(r,r^{\prime}) + \ldots.
\end{equation}
Because of translation invariance in the absence of background fields, the zeroth-order Green's
function depends only on the difference $r-r^{\prime}$. The same is not true, in general, for the first-order
part of the Green's function. The Fourier transform of $G^{(0)}(r-r^{\prime})$ follows directly from the model
Hamiltonian (\ref{lattice-d-def-be}), i.e.,
\begin{equation}
\label{lattice-G0-def-be}
G^{(0)}(\omega;\mathbf{k}) =  \frac{i\left[\omega+\mu
+(\mathbf{d}\cdot\bm{\sigma})\right]}{[\omega+\mu+i0\sign{\omega}]^2-|\mathbf{d}|^2} ,
\end{equation}
where we introduced a nonzero chemical potential $\mu$ and omitted $d_0$. Now, the
correction to the Green's function linear in the electromagnetic field can be
obtained by using a perturbative expansion in the interaction Hamiltonian, i.e.,
\begin{equation}
G^{(1)}(r,r^{\prime})=-i\int dr^{\prime \prime}G^{(0)}(r-r^{\prime\prime})\mathcal{H}_{\rm int}(r^{\prime\prime})
G^{(0)}(r^{\prime \prime}-r^{\prime}).
\label{first-order-correction}
\end{equation}
In the next two sections, we will use the above representation for the Green's function
in order to study the linear response of Weyl semimetals to background electromagnetic fields.
A similar representation, although with a different interaction Hamiltonian, will be also used later
in the case of strain-induced pseudoelectromagnetic fields.

\section{Response to background magnetic field}
\label{sec:lattice-first}

In this section we derive explicit expressions for the electric charge and current densities
to the linear order in a background magnetic field. We assume that the field points in the $+z$ direction
and is described by the vector potential in the Landau gauge, i.e., $\mathbf{A}=\left(0,xB,0\right)$.
By making use of the definitions in Eqs.~(\ref{lattice-rho-def}) and (\ref{lattice-J-def}), as well
as the linear-order correction to the Green's function $G^{(1)}(r,r^{\prime})$ obtained in
Appendix~\ref{Sec:App-Green-first}, we find
\begin{eqnarray}
\label{lattice-first-rho}
\rho&=& \frac{eB}{2} \int \frac{d\omega d^3\mathbf{k}}{(2\pi)^4} \mbox{tr}\Bigg\{\left[\partial_{k_x}G^{(0)}(\omega; \mathbf{k})\right]
j_{y}(\mathbf{k})
G^{(0)}(\omega;\mathbf{k}) - G^{(0)}(\omega; \mathbf{k})j_{y}(\mathbf{k}) \left[\partial_{k_x}G^{(0)}(\omega; \mathbf{k})\right] \Bigg\},\\
\label{lattice-first-J}
J_{n}&=& \frac{B}{2} \int \frac{d\omega d^3\mathbf{k}}{(2\pi)^4} \mbox{tr}\Bigg\{j_{n}(\mathbf{k})\left[\partial_{k_x}
G^{(0)}(\omega,\mathbf{k})\right]
j_{y}(\mathbf{k}) G^{(0)}(\omega;\mathbf{k})
- j_{n}(\mathbf{k})G^{(0)}(\omega; \mathbf{k})j_{y}(\mathbf{k}) \left[\partial_{k_x}G^{(0)}(\omega; \mathbf{k})\right] \nonumber\\
&+&\delta_{n,x} \left[\partial_{k_x}j_{x}(\mathbf{k})\right]G^{(0)}(\omega; \mathbf{k})j_{y}(\mathbf{k}) G^{(0)}(\omega; \mathbf{k}) +2ix\delta_{n,x} j_{x}(\mathbf{k})G^{(0)}(\omega; \mathbf{k})j_y(\mathbf{k}) G^{(0)}(\omega; \mathbf{k})\Bigg\}.
\end{eqnarray}
Note that both densities are given in terms of the zeroth-order Green's function $G^{(0)}(\omega;\mathbf{k})$ defined
in Eq.~(\ref{lattice-G0-def-be}), as well as its derivatives with respect to the quasimomentum.

Let us first calculate the electric charge density (\ref{lattice-first-rho}). After
substituting the explicit form of the zeroth-order Green's function $G^{(0)}(\omega;\mathbf{k})$, we find
that the integration over $\omega$ can be performed analytically.
For the details of the derivation, see Appendix~\ref{sec:App-key-point-charge-current-B}.
In the case of the vanishing chemical potential, the final result reads
\begin{equation}
\rho = -\frac{e^2B}{2(2\pi)^3} \int d^3\mathbf{k}\, \left(\hat{\mathbf{d}}\cdot\Big[(\partial_{k_x}\hat{\mathbf{d}})
\times(\partial_{k_y}\hat{\mathbf{d}})\Big]\right),
\label{lattice-topology-rho-1}
\end{equation}
where $\hat{\mathbf{d}}\equiv\mathbf{d}/|\mathbf{d}|$. Its topological nature is evident
from the fact that the integrand is proportional to the $z$ component
of the Berry curvature. Indeed, the latter is defined by \cite{Haldane}
\begin{equation}
\Omega_{i}=\sum_{l,m=1}^3\frac{\epsilon_{i l m}}{4} \left(\hat{\mathbf{d}}\cdot\Big[(\partial_{k_{l}}\hat{\mathbf{d}})
\times(\partial_{k_{m}}\hat{\mathbf{d}})\Big]\right).
\label{lattice-topology-inv-Berry}
\end{equation}
A component of the Berry curvature can be also viewed as the Jacobian of the mapping of a two-dimensional section
of the Brillouin zone onto the unit sphere, i.e., $T^2\to S^2$.
When integrated over the area of the cross section (for example, the $k_x$-$k_y$ plane), it counts the winding
number of the mapping or the Chern number \cite{Bernevig:2013}
\begin{equation}
\mathcal{C}(k_z)=\frac{1}{4\pi} \int dk_x\,dk_y\, \left(\hat{\mathbf{d}}\cdot\Big[(\partial_{k_x}\hat{\mathbf{d}})
\times(\partial_{k_y}\hat{\mathbf{d}})\Big]\right).
\label{lattice-topology-inv-1}
\end{equation}
As is easy to check, $\mathcal{C}(k_z)$ depends on $k_z$ and vanishes for $|k_z| \ge b_z$ in the model under
consideration. By integrating the Chern number over $k_z$, we find that the result for $\rho$
coincides with the topological Bardeen-Zumino expression for the electric charge density induced
by a magnetic field \cite{Bardeen,Landsteiner:2013sja,Landsteiner:2016}, generalized to the case of a multi-Weyl semimetal, i.e.,
\begin{equation}
\rho_{\text{{\tiny BZ}}}=-n\frac{e^2B b_z}{2\pi^2}.
\label{consistent-charge-density}
\end{equation}
Here $n$ denotes the topological charge of the Weyl nodes in multi-Weyl semimetals, i.e., $n=1$ in a Weyl semimetal, $n=2$ in a double-Weyl semimetal, and $n=3$ in a triple-one. Indeed, as we show in Appendix~\ref{Sec:App-multi-Weyl}, the corresponding lattice
models of multi-Weyl semimetals can be defined by the same Hamiltonian as in Eq.~(\ref{lattice-d-def-be}) but with a different choice
of functions $\mathbf{d}$. Then, the electric charge density at $\mu=0$ will be formally
given by the topological expression (\ref{lattice-topology-rho-1}) proportional
to the winding number.
This finding confirms that the Bardeen-Zumino contribution to the electric charge
density is reproduced exactly in lattice models of multi-Weyl semimetals with finite Brillouin zones.

As we already mentioned in the Introduction, the result in Eq.~(\ref{consistent-charge-density})
cannot be captured by the chiral kinetic theory. The easiest way to see this is to note that the
equations of the chiral kinetic theory do not contain the chiral shift parameter
$\mathbf{b}$ at all. It is rather interesting, as we argue below, that the topological
Bardeen-Zumino term appears to be the only contribution that the chiral kinetic theory fails to reproduce.
In order to fully substantiate this claim, it is instructive to consider the calculation of the electric charge
density in the same lattice model at nonzero chemical potential $\mu$.

To the linear order in magnetic field $\mathbf{B}$, the complete expression for the electric charge density
at nonzero chemical potential $\mu$ reads
\begin{equation}
\rho =\rho_{\text{{\tiny BZ}}} +\rho_{\mu},
\label{lattice-topology-mu-rho}
\end{equation}
where the additional ``matter" part of the density is given by
\begin{equation}
\rho_{\mu} = \frac{e^2B}{2(2\pi)^3} \int \frac{d^3\mathbf{k} }{|\mathbf{d}|^3}\left(\mathbf{d}\cdot\left[(\partial_{k_x}\mathbf{d})
\times(\partial_{k_y}\mathbf{d})
\right]\right) \theta\left(|\mu|-|\mathbf{d}|\right) + \frac{e^2B}{(2\pi)^3} \int \frac{d^3\mathbf{k} }{|\mathbf{d}|}
\delta\left(\mu^2-|\mathbf{d}|^2\right) \left(\mathbf{d}\cdot\left[(\partial_{k_x}\mathbf{d})
\times(\partial_{k_y}\mathbf{d})\right]\right).
\label{lattice-rho-mu-only}
\end{equation}
By making use of the Berry curvature defined in Eq.~(\ref{lattice-topology-inv-Berry}), the matter part can be
cast in a simpler form, i.e.,
\begin{equation}
\rho_{\mu} = e^2 \int \frac{d^3\mathbf{k}}{(2\pi)^3}\, (\mathbf{B}\cdot\bm{\Omega}) \left[
\theta\left(|\mu|-|\mathbf{d}|\right) +|\mathbf{d}|\,\delta\left(|\mu|-|\mathbf{d}|\right) \right].
\label{lattice-topology-mu-rho-01}
\end{equation}
For a specific set of model parameters in Appendix~\ref{Sec:App-model}, it is straightforward to
calculate the corresponding contribution to the charge density using numerical methods. It is
much more instructive, however, to compare Eq.~(\ref{lattice-topology-mu-rho-01}) with its
counterpart in the chiral kinetic theory (see, e.g., Ref.~\cite{Gorbar:2017cwv}):
\begin{eqnarray}
\rho_{\text{{\tiny CKT}}} &=& \sum_{\eta=\pm} \eta e\int \frac{d^3\mathbf{k}}{(2\pi)^3} \left[1+\eta e(\mathbf{B}\cdot\bm{\Omega})\right] \left[1+e^{\left(v_Fk-e \eta v_Fk (\mathbf{B}\cdot\bm{\Omega})  -\eta\mu\right)/T}\right] -\sum_{\eta=\pm} \eta e\int \frac{d^3\mathbf{k}}{(2\pi)^3} n_{\eta}(v_Fk)\nonumber\\
&\simeq& \sum_{\eta=\pm} \eta e\int \frac{d^3\mathbf{k}}{(2\pi)^3} \left[n_{\eta}(v_Fk) +\eta e(\mathbf{B}\cdot\bm{\Omega})n_{\eta}(v_Fk) -\eta v_Fk e(\mathbf{B}\cdot\bm{\Omega})n_{\eta}^{\prime}(v_Fk)\right] -\sum_{\eta=\pm} \eta e\int \frac{d^3\mathbf{k}}{(2\pi)^3} n_{\eta}(v_Fk),
\end{eqnarray}
where $n_{\eta}(x)=1/\left[1+e^{\left(x -\eta\mu\right)/T}\right]$ is the Fermi-Dirac distribution and
we set $c=1$ according to the conventions in this paper. In the zero temperature limit, the corresponding
contribution linear in the magnetic field reads
\begin{equation}
\rho_{\text{{\tiny CKT}}} =e^2 \int \frac{d^3\mathbf{k}}{(2\pi)^3}\,(\mathbf{B}\cdot\bm{\Omega}) \left[ \theta(|\mu|-v_Fk)
+v_Fk \delta(|\mu|-v_Fk)\right].
\label{rho-CKT}
\end{equation}
As we see, the chiral kinetic theory result in Eq.~(\ref{rho-CKT}) reproduces exactly the matter
part of the charge density in Eq.~(\ref{lattice-topology-mu-rho-01}) obtained in the lattice
model of a Weyl semimetal if we set $\mathbf{d}=v_F\mathbf{k}$.

Let us discuss now the electric current density. As in the case of the electric charge density, after
substituting the zeroth-order Green's function $G^{(0)}(\omega; \mathbf{k})$, see Eq.~(\ref{lattice-G0-def-be}),
into the expression for the current density (\ref{lattice-first-J}) and performing the integration
over $\omega$ analytically (see Appendix~\ref{sec:App-key-point-charge-current-B} for details), we obtain
\begin{equation}
J_n=- e^2B\int \frac{d^3\mathbf{k}}{(2\pi)^3} \sign{\mu} \sum_{i,l,m=1}^{3}\epsilon_{ilm}
(\partial_{k_n}d_{i})(\partial_{k_x}d_{l})(\partial_{k_y}d_{m}) \delta\left(\mu^2-|\mathbf{d}|^2\right),
\label{lattice-topology-J-n-1}
\end{equation}
where we omitted the imaginary, as well as coordinate-dependent terms that vanish after the
integration over the momentum. Further, after integrating over the whole Brillouin zone, we find
that the electric current density (\ref{lattice-topology-J-n-1}) linear in a constant background magnetic
field vanishes. This means that the chiral magnetic effect is absent in the equilibrium state,
which is in agreement with general requirements of the band theory of solids \cite{Franz:2013,Aschcroft}.

\section{Response to background electric field}
\label{sec:Kubo}

In order to identify the topological Bardeen-Zumino contributions in the electric current density, one needs to
study the response of the lattice Weyl model to a background electric field. The corresponding analysis
is performed in this section using the Kubo's linear response theory.

In the framework of the Kubo's theory, the charge and current densities can be written in the form
$\rho = \sigma_{0m} E_m$ and $J_n = \sigma_{nm} E_m$, respectively. Here $\sigma_{\nu m}$
with $\nu=(0,x,y,z)$ describes the response of the charge and direct current densities to the electric field, which is related to the
polarization tensor $\Pi_{\nu m}(\Omega+i0;\mathbf{0})$ via the standard relation:
\begin{equation}
\sigma_{\nu m} = -\lim_{\Omega\to0} \frac{i}{\Omega} \Pi_{\nu m}(\Omega+i0;\mathbf{0}),
\label{Kubo-sigma-ij-def}
\end{equation}
where the polarization tensor is defined in terms of the quasiparticle Green's function, i.e.,
\begin{equation}
\Pi_{\nu m}(\Omega+i0;\mathbf{0}) = -T\sum_{l=-\infty}^{\infty} \int \frac{d^3\mathbf{k}}{(2\pi)^3}
\mbox{tr}\left[j_{\nu}(\mathbf{k})
G^{(0)}(i\omega_l;\mathbf{k})j_{m}(\mathbf{k})G^{(0)}(i\omega_l-\Omega;\mathbf{k})\right]
\label{Kubo-Pi-ij-def}
\end{equation}
and $\omega_{l}=(2l+1)\pi T$ (with $l\in\mathbb{Z}$) are the fermion Matsubara frequencies.
Here, we also introduced the four-vector $j_{\nu}=\left(e, -\mathbf{j}(\mathbf{k})\right)$. By following the standard
approach, it is convenient to rewrite the Green's function in terms of its spectral function
\begin{equation}
G^{(0)}(i\omega_l; \mathbf{k})=\int_{-\infty}^{\infty} d\omega \frac{A(\omega; \mathbf{k})}{i\omega_l+\mu-\omega},
\label{Kubo-Green-spectral-def}
\end{equation}
where, by definition,
\begin{equation}
A(\omega; \mathbf{k}) \equiv \frac{i}{2\pi}
\left[G^{(0)}(\omega+i0; \mathbf{k})-G^{(0)}(\omega-i0; \mathbf{k})\right]_{\mu=0} =
i\sum_{s=\pm}\frac{|\mathbf{d}|+s(\mathbf{d}\cdot\bm{\sigma})}{2|\mathbf{d}|} \delta\left(\omega-s|\mathbf{d}|\right).
\label{Kubo-spectral-function-def}
\end{equation}
As indicated by the delta function, this spectral function describes noninteracting quasiparticles with
a vanishing decay width. In realistic models, of course, the corresponding decay width should be nonzero.
This can be implemented by using a phenomenological model, in which the delta function is replaced with the
Lorentzian distribution, i.e.,
\begin{equation}
\delta_{\Gamma}(\omega-s|\mathbf{d}|)\equiv \frac{1}{\pi} \frac{\Gamma(\omega)}{(\omega-s|\mathbf{d}|)^2+\Gamma^2(\omega)}.
\label{Kubo-d-Gamma}
\end{equation}
Note that, at low energies, the quasiparticle width includes a constant part $\Gamma_0$ as well
as a frequency-dependent part proportional to $\omega^2$ \cite{Burkov:2011}, i.e., $\Gamma(\omega)=\Gamma_0(1+\omega^2/\epsilon_0^2)$. Henceforth, we will omit the argument of $\Gamma$.
By making use of the spectral representation for the Green's function, Eq.~(\ref{Kubo-sigma-ij-def}) can be recast in the following form:
\begin{equation}
\sigma_{\nu m} = \lim_{\Omega\to0}\frac{i}{\Omega} T \sum_{l=-\infty}^{\infty} \int\frac{d^3\mathbf{k}}{(2\pi)^3}
\int \int d\omega d\omega^{\prime} \frac{\mbox{tr}\left[j_{\nu}(\mathbf{k})A(\omega; \mathbf{k})j_{m}(\mathbf{k})
A(\omega^{\prime}; \mathbf{k})\right]}{\left(i\omega_l+\mu-\omega\right)\left(i\omega_l-\Omega-i0+\mu-\omega^{\prime}\right)}.
\label{Kubo-conductivity-calc-1}
\end{equation}
After performing the summation over the Matsubara frequencies and setting $T=0$,
we derive the following results:
\begin{eqnarray}
\label{Kubo-conductivity-calc-rho}
\sigma_{0m} &=& e^2\pi \int\frac{d^3\mathbf{k}}{(2\pi)^3} \frac{\delta_{\Gamma}^2(\mu-|\mathbf{d}|) -\delta_{\Gamma}^2(\mu+|\mathbf{d}|)}{|\mathbf{d}|}
\left(\mathbf{d}\cdot(\partial_{k_m}\mathbf{d})\right),\\
\sigma_{nn} &=&e^2 \pi \int\frac{d^3\mathbf{k}}{(2\pi)^3} \frac{1}{|\mathbf{d}|^2}\Big\{\left((\partial_{k_n}\mathbf{d})\cdot\mathbf{d}\right)^2 \left[\delta_{\Gamma}^2(\mu-|\mathbf{d}|)+\delta_{\Gamma}^2(\mu+|\mathbf{d}|)\right]\nonumber\\
&-&2\delta_{\Gamma}(\mu-|\mathbf{d}|)\delta_{\Gamma}(\mu+|\mathbf{d}|)\left[\left((\partial_{k_n}\mathbf{d})\cdot\mathbf{d}\right)^2 -|\mathbf{d}|^2(\partial_{k_n}\mathbf{d})^2
%\cdot\partial_{k_n}\mathbf{d})
\right]
\Big\}.
\label{Kubo-conductivity-calc-Re}
\end{eqnarray}
In addition, there are also nonzero off-diagonal components of the conductivity tensor. In the clean limit,
in particular, the latter read
\begin{equation}
\sigma_{nm} = -\frac{e^2}{2}\int\frac{d^3\mathbf{k}}{(2\pi)^3}
 \left(\hat{\mathbf{d}}\cdot\left[(\partial_{k_n}\hat{\mathbf{d}})\times(\partial_{k_m}\hat{\mathbf{d}})\right]\right) \left[1-\theta(|\mu|-|\mathbf{d}|)
 \right],
\label{Kubo-conductivity-calc-Im-Gamma0}
\end{equation}
and the only nonvanishing components of this conductivity tensor are $\sigma_{12}=-\sigma_{21}$.

\begin{figure}[!ht]
\begin{center}
\includegraphics[width=0.45\textwidth]{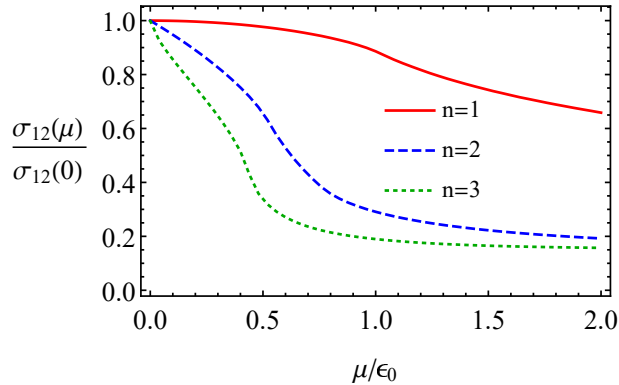}
\caption{The dependence of the zero-temperature Hall conductivity $\sigma_{12}$ in a Weyl semimetal (red solid line), a double-Weyl semimetal (blue dashed line), and a triple- Weyl semimetal (green dotted line) on
the chemical potential. Here $\epsilon_0$ is the value of energy (\ref{lattice-E}) at $\mathbf{k}=\mathbf{0}$.
The numerical results are calculated for the model parameters in
Appendix~\ref{Sec:App-model}.}
\label{fig:lattice-Kubo}
\end{center}
\end{figure}

It should be clear that the off-diagonal conductivity $\sigma_{12}$ in
Eq.~(\ref{Kubo-conductivity-calc-Im-Gamma0}) has a topological origin in the limit of the
vanishing chemical potential ($\mu\to 0$). Indeed, as in the case of the charge density
(\ref{lattice-topology-rho-1}), it is determined by the integral of the Chern number
(\ref{lattice-topology-inv-1}), or the winding number, which is equal to the topological charge of the Weyl nodes $n$. After calculating the corresponding
integral, we derive the following explicit result for the anomalous Hall conductivity:
\begin{equation}
\sigma_{12}=-\sigma_{21} \stackrel{\mu=0}{=} -n\frac{e^2 b_z}{2\pi^2}.
\label{Kubo-conductivity-12}
\end{equation}
This result corresponds to the expected topological Bardeen-Zumino current that describes
the anomalous Hall conductivity \cite{Ran,Burkov:2011ene,Grushin-AHE,Zyuzin,Goswami}.
For a set of model parameters in Appendix~\ref{Sec:App-model}, the zero-temperature Hall
conductivity $\sigma_{12}$ as a function of the chemical potential $\mu$ is plotted in
Fig.~\ref{fig:lattice-Kubo}. From a physics viewpoint, the corresponding result includes
both the anomalous and matter contributions. It is interesting to note that the total
Hall conductivity decreases with increasing the absolute value of chemical potential. Moreover, the corresponding dependence is much steeper in the case of double- and triple-Weyl semimetals.
Since the second term with the theta-function in Eq.~(\ref{Kubo-conductivity-calc-Im-Gamma0}) contains the integration over all filled quasiparticle states, naively it contradicts the conventional wisdom of the Fermi liquid theory which states that the conductivity is related only to the states on the Fermi surface. However, it was shown in Ref.~\cite{Haldane} that the corresponding term can be rewritten as a Fermi surface integral and is always present in realistic models at nonzero chemical potential. According to Ref.~\cite{Kohmoto:1992}, the off-diagonal conductivity of a completely filled or empty band should be proportional to a primitive reciprocal lattice vector (including zero). Notice that in the model under consideration this requirement is trivially satisfied, because $\sigma_{12}$ tends to zero when the bands are completely filled or empty ($\mu\to\pm\infty$).

%%%%%%%%%%%%%%%%%%
%%%%%%%%%%%%%%%%%%

\section{Response to a strain-induced pseudomagnetic field}
\label{sec:lattice-B5}

In the preceding two sections, we studied the electric charge and current response
of a Weyl semimetal to external magnetic and electric fields in the lattice model (\ref{lattice-d-def-be}).
It is also of interest to investigate the response of Weyl materials to pseudoelectromagnetic fields.
The latter, as we mentioned in the Introduction, could be generated by applying mechanical
deformations to Weyl semimetals. In this section, we consider the response of a Weyl semimetal with $n=1$ to a strain-induced
pseudomagnetic field $\mathbf{B}_5$. The response to a pseudoelectric field $\mathbf{E}_5$
will be studied in the next section.

According to Ref.~\cite{Pikulin:2016}, strains in Weyl materials lead to the following additional
terms in Hamiltonian (\ref{lattice-d-def-be}):
\begin{equation}
\label{lattice-B5-dh}
\delta h_{\rm strain}= \Lambda \left(u_{13}\sigma_x +u_{23}\sigma_y\right)\sin{(ak_z)}
-t_1 u_{33} \sigma_z \cos{(ak_z)},
\end{equation}
where $u_{ij}=\left(\partial_{i}u_j+\partial_{j}u_i\right)/2$ is the symmetrized strain tensor and
$\mathbf{u}=(u_1,u_2,u_3)$ is the displacement vector. In the vicinity of Weyl nodes,
the additional terms given by Eq.~(\ref{lattice-B5-dh}) can be interpreted as the interaction
Hamiltonian of Weyl quasiparticles with the background axial gauge field
\begin{equation}
\label{lattice-B5-Franz-gauge}
\mathbf{A}_5=  \frac{1}{ea}\left[u_{13}\sin{(ab_z)},u_{23}\sin{(ab_z)},u_{33}\cot{(ab_z)}\right].
\end{equation}
As is clear, not all strains in Weyl materials can produce nontrivial pseudoelectromagnetic fields. For example,
a time independent $u_{33}$ describes a stretching of the crystal along the $+z$ direction. This leads to a simple redefinition of the parameter $t_1\to(1-u_{33})t_1$ that,
in turn, modifies the value of the chiral shift parameter (\ref{lattice-bz}). In the rest of this section,
we will primarily concentrate on the case of static strains with $\bm{\nabla}\times\mathbf{A}_5 \neq \mathbf{0}$
that describe pseudomagnetic fields.

\begin{figure}[!ht]
\begin{center}
\includegraphics[width=0.32\textwidth]{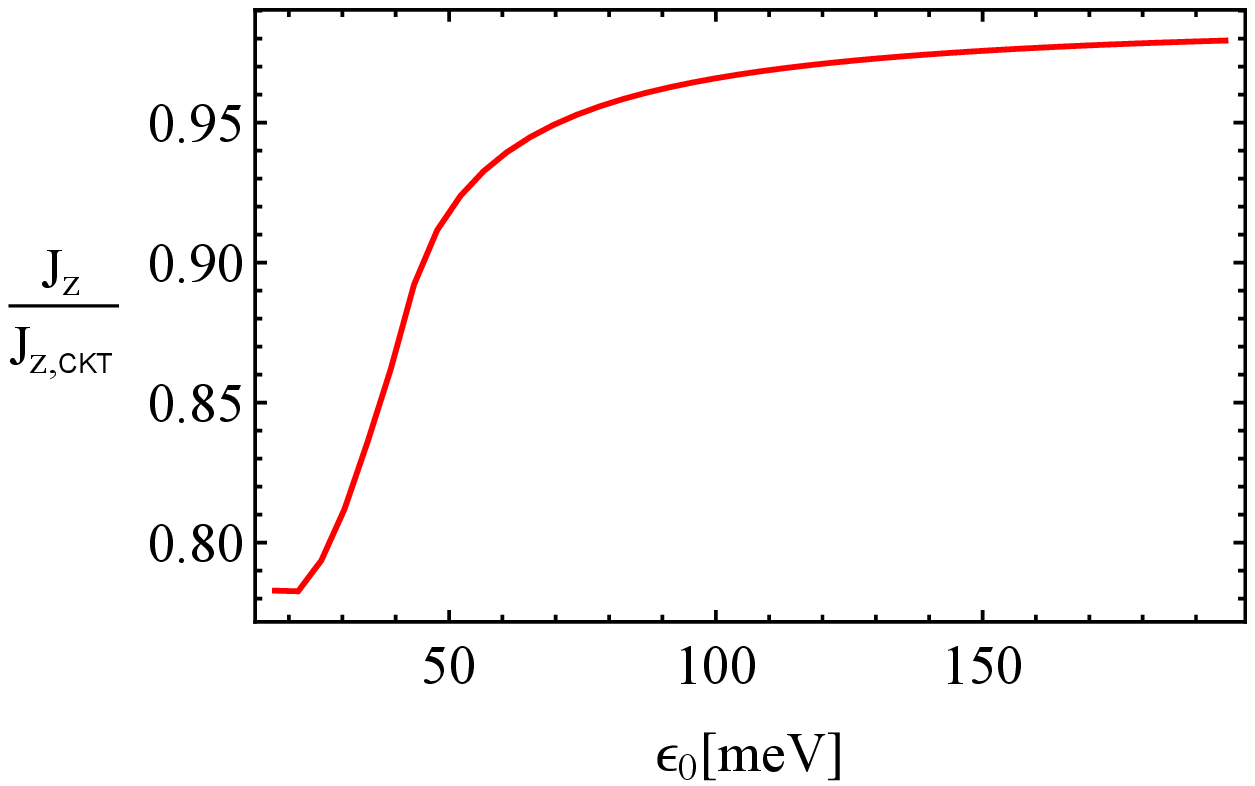}\hfill
\includegraphics[width=0.32\textwidth]{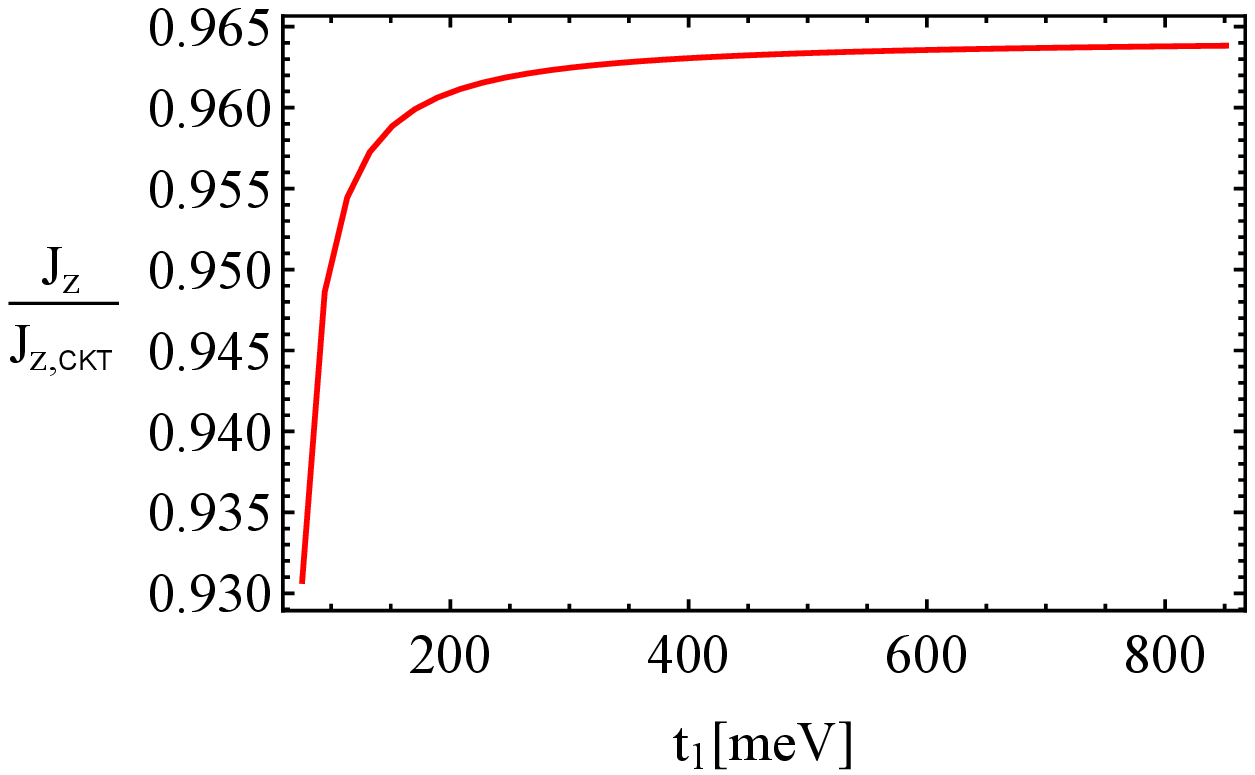}\hfill
\includegraphics[width=0.32\textwidth]{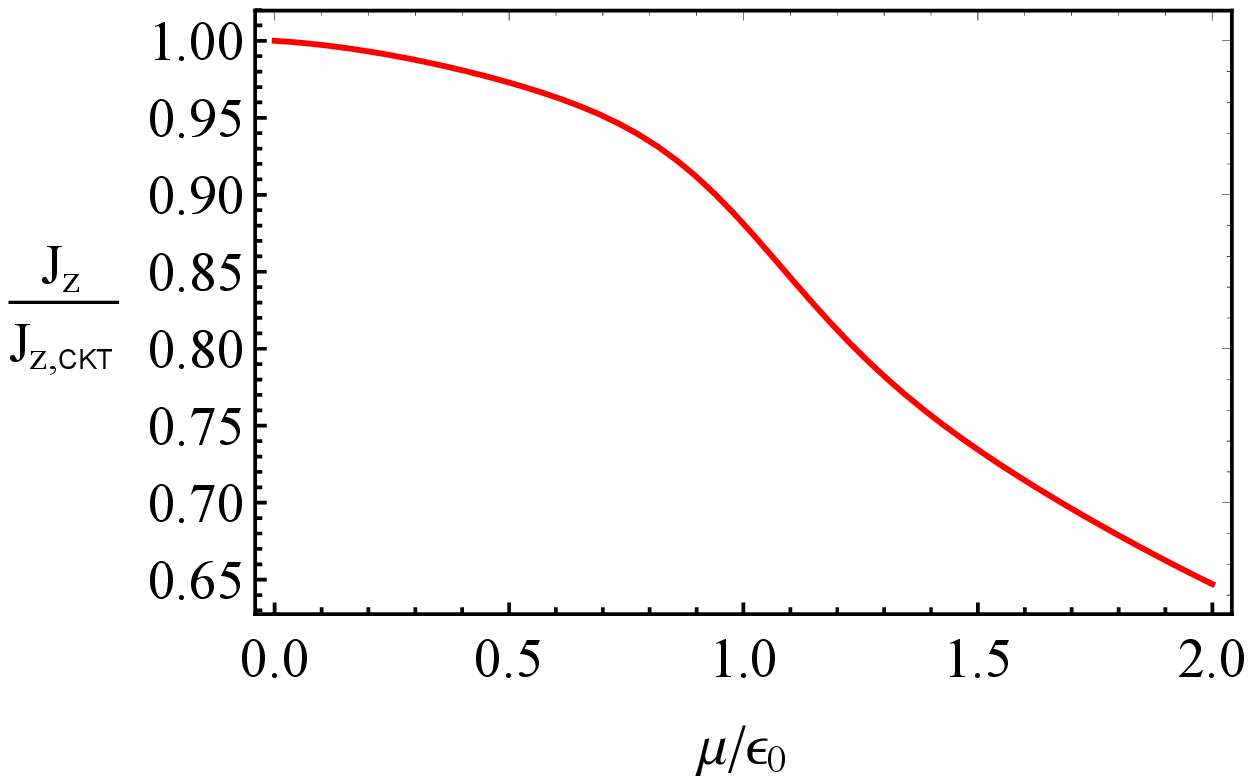}
\caption{The electric current density $J_z$ in the direction of the
pseudomagnetic field as a function of $\epsilon_0$ (left panel), $t_1$ (middle panel), and
$\mu/\epsilon_0$ (right panel). In the left and middle panels, the chemical potential is
$\mu=43.45~\mbox{meV}$. In the middle and right panels, we used $\epsilon_0=86.9~\mbox{meV}$,
which corresponds to the ``dome" energy at $k=0$ in the model defined in Appendix~\ref{Sec:App-model}.}
\label{fig:lattice-B5-J3-mu}
\end{center}
\end{figure}

We consider the case of a constant strain-induced pseudomagnetic field
along the $z$ direction. Such a field can be induced, for example, by applying torsion to a wire
made of a Weyl material,
\begin{equation}
\label{lattice-B5-Franz-B5}
\mathbf{B}_5\equiv \bm{\nabla}\times\mathbf{A}_5= -\frac{\theta}{L e a} \sin{(ab_z)} \hat{\mathbf{z}},
\end{equation}
where $\mathbf{u}= \theta z [\mathbf{r}\times\hat{\mathbf{z}}]/L$,
 $\theta$ is the torsion angle, and $L$ is the length of the crystal.
Then the interaction Hamiltonian (\ref{lattice-B5-dh}) takes the following explicit form:
\begin{equation}
\label{lattice-B5-dh-1}
\delta h_{\rm strain}=
\frac{eB_5 a \Lambda \sin{(ak_z)} }{2\sin{(ab_z)}} \left( x \sigma_y - y \sigma_x \right).
\end{equation}
It should be noted that the latter has the same structure as the interaction Hamiltonian in the
case of a constant magnetic field $\mathbf{B}$. This becomes evident by introducing
the following notation for the transverse components of the axial current:
\begin{eqnarray}
\label{lattice-B5-j5-1}
j^5_x &=& \frac{ea \Lambda}{\sin{(ab_z)}}\sigma_x \sin{(ak_z)},\\
j^5_y &=& \frac{ea \Lambda}{\sin{(ab_z)}}\sigma_y \sin{(ak_z)}.
\label{lattice-B5-j5-1-ee}
\end{eqnarray}
Then, in the full analogy with Eqs.~(\ref{lattice-first-rho}) and (\ref{lattice-first-J}), we derive
the following expressions for the electric charge and current densities induced by the pseudomagnetic
field:
\begin{eqnarray}
\label{lattice-B5-rho}
\rho&=& \frac{eB_{5}}{4} \int \frac{d\omega d^3\mathbf{k}}{(2\pi)^4} \mbox{tr}\Bigg[ -(\partial_{k_y}G^{(0)})j^5_x G^{(0)} + G^{(0)}j^5_x (\partial_{k_y}G^{(0)})
+(\partial_{k_x}G^{(0)})j^5_y G^{(0)} - G^{(0)}j^5_y (\partial_{k_x}G^{(0)}) \Bigg], \\
\label{lattice-B5-J}
J_{n}&=& \frac{B_5}{4} \int \frac{d\omega d^3\mathbf{k}}{(2\pi)^4} \mbox{tr}\Bigg[ -j_{n}(\partial_{k_y}G^{(0)})j^5_x G^{(0)} + j_{n}G^{(0)}j^5_x (\partial_{k_y}G^{(0)})
-\delta_{n,y} (\partial_{k_y}j_{y})G^{(0)}j^5_x G^{(0)} -2iy\delta_{n,y} j_{y}G^{(0)}j^5_x G^{(0)}\nonumber\\
&+&j_{n}(\partial_{k_x}G^{(0)})j^5_y G^{(0)} - j_{n}G^{(0)}j^5_y (\partial_{k_x}G^{(0)})
+\delta_{n,x} (\partial_{k_x}j_{x})G^{(0)} j^5_y G^{(0)} +2ix\delta_{n,x} j_{x}G^{(0)} j^5_y G^{(0)} \Bigg],
\end{eqnarray}
where for the sake of simplicity we dropped the arguments of $\mathbf{j}$, $\mathbf{j}^5$, and $G^{(0)}$.
By making use of the integrals in Eqs.~(\ref{mu-omega-int-be})--(\ref{mu-omega-int-ee}),
we can integrate over $\omega$. Then, by setting $u_{33}=0$
(see also Appendix~\ref{sec:App-key-point-charge-current-B}), we obtain
\begin{eqnarray}
\label{lattice-B5-rho-BZ}
\rho_{\rm top}&=& \frac{e^2 B_5 a \Lambda}{4\sin{(ab_z)}}  \int \frac{d^3\mathbf{k}}{(2\pi)^3}  \frac{\sin{(ak_z)}}{|\mathbf{d}|^3}\left\{[(\partial_{k_y}\mathbf{d})
\times\mathbf{d}]_x -[(\partial_{k_x}\mathbf{d})\times\mathbf{d}]_y \right\},\\
\label{lattice-B5-rho-mu}
\rho_{\mu}&=& -\frac{e^2 B_5 a \Lambda}{4\sin{(ab_z)}}  \int \frac{d^3\mathbf{k}}{(2\pi)^3}  \frac{\sin{(ak_z)}}{|\mathbf{d}|^3} \left\{[(\partial_{k_y}\mathbf{d})
\times\mathbf{d}]_x -[(\partial_{k_x}\mathbf{d})\times\mathbf{d}]_y \right\} \left[\theta\left(|\mu|-|\mathbf{d}|\right)+|\mathbf{d}|\delta\left(|\mu|-|\mathbf{d}|\right) \right],
\end{eqnarray}
and
\begin{equation}
J_n= -\frac{e^2 B_5 a \Lambda}{2\sin{(ab_z)}} \int \frac{d^3\mathbf{k}}{(2\pi)^3} \sin{(ak_z)} \sign{\mu}
\delta\left(\mu^2-|\mathbf{d}|^2\right)\left\{\left[(\partial_{k_n}\mathbf{d})\times(\partial_{k_y}\mathbf{d})\right]_x
-\left[(\partial_{k_n}\mathbf{d})\times(\partial_{k_x}\mathbf{d})\right]_y\right\},
\label{lattice-B5-J-n-mu}
\end{equation}
where we also omitted imaginary and coordinate dependent terms which vanish after the integration over the
whole Brillouin zone.
Our numerical calculations show that $\rho_{\rm top}=\rho_{\mu}=J_{x}=J_y=0$. Thus, unlike the magnetic field
considered in Sec.~\ref{sec:lattice-first}, the pseudomagnetic field does not induce any electric charge density.
On the other hand, the component of electric current along the direction of the pseudomagnetic fields is nonzero.
The dependencies of $J_z$ for a Weyl semimetal on the parameters $\epsilon_0$, $t_1$, and chemical potential $\mu$ are shown
in the left, middle, and right panels of Fig.~\ref{fig:lattice-B5-J3-mu}, respectively. In particular, the right panel
of  of Fig.~\ref{fig:lattice-B5-J3-mu} shows that, at sufficiently small values of $\mu$, the electric current agrees
with the corresponding expression in the chiral kinetic theory \cite{Zhou:2012ix,Grushin-Vishwanath:2016},
\begin{equation}
J_{z,\text{{\tiny CKT}}}= -\frac{e^2 \mu B_5}{2\pi^2}.
\label{lattice-B5-Jz-top}
\end{equation}
Notably, however, the latter is not exact and receives corrections at large enough values of $\mu$.
Also, as we see from the left and middle panels of Fig.~\ref{fig:lattice-B5-J3-mu}, the result for the
current $J_z$ depends on other model parameters and, consequently, is not fully protected
by topology. While this might appear surprising, the reason for this is rather simple and related to the
fact that the interpretation of the strain-induced background field (\ref{lattice-B5-Franz-gauge}) as
a conventional axial vector potential $\mathbf{A}_5$ deteriorates outside of the immediate
vicinity of the Weyl nodes.

%%%%%%%%%%%%%%%%%%
%%%%%%%%%%%%%%%%%%

\section{Response to a strain-induced pseudoelectric field}
\label{sec:Kubo-E5}

In this section, we study the response to a strain-induced pseudoelectric field $\mathbf{E}_5$.
Such a field can be generated by time-dependent deformations of a Weyl crystal.
As in Sec.~\ref{sec:Kubo}, here we use the Kubo's linear response theory.
By reexpressing the deformation tensor components $u_{ij}\sim t$ in terms of the axial
vector potential $\mathbf{A}_5$ and using the relation $\mathbf{E}_5=-\partial_{t}\mathbf{A}_5$,
we obtain the following interaction Hamiltonian:
\begin{equation}
\label{Kubo-E5-dh-be}
\delta h_{\rm strain}=
 -\frac{eE_{5,x} a\Lambda \sin{(ak_z)}}{\sin{(ab_z)}}\sigma_x t
 -\frac{eE_{5,y} a\Lambda \sin{(ak_z)}}{\sin{(ab_z)}}\sigma_y t
 + \frac{eE_{5,z} at_1 \cos{(ak_z)}}{\cot{(ab_z)}} \sigma_z t.
\end{equation}
By comparing this with the electromagnetic interaction Hamiltonian
(\ref{lattice-H-int}), we found that the components of the current density operator $j_x^5$ and $j_y^5$ are given by Eqs.~(\ref{lattice-B5-j5-1}) and (\ref{lattice-B5-j5-1-ee}), respectively, and $j_z^5$ reads
\begin{equation}
j_z^5 = -e\frac{at_1 \cos{(ak_z)}}{\cot{(ab_z)}}\sigma_z.
\label{Kubo-E5-j-ee}
\end{equation}

The DC conductivity tensor $\sigma_{\nu m}^{(5)}$ that quantifies
the response of the \emph{electric} charge and current densities to a background
\emph{pseudoelectric} field $\mathbf{E}_5$ is given by the standard Kubo's formula,
\begin{equation}
\sigma_{\nu m}^{(5)} = -\lim_{\Omega\to0} \frac{i}{\Omega} \Pi_{\nu m}^{(5)}(\Omega+i0;\mathbf{0}),
\label{Kubo-E5-sigma-ij-def}
\end{equation}
where the current-current correlator on the right-hand side is defined in terms of the
quasiparticle Green's function as follows:
\begin{equation}
\Pi_{\nu m}^{(5)}(\Omega+i0;\mathbf{0}) = -T\sum_{l=-\infty}^{\infty} \int \frac{d^3\mathbf{k}}{(2\pi)^3} \mbox{tr}\left[j_{\nu}(\mathbf{k})
G^{(0)}(i\omega_l;\mathbf{k})j_{m}^{5}(\mathbf{k})G^{(0)}(i\omega_l-\Omega;\mathbf{k})\right].
\label{Kubo-E5-Pi-ij-def}
\end{equation}
By following the same method as in Sec.~\ref{sec:Kubo}, we can express the conductivity
in terms of the spectral function,
\begin{equation}
\sigma_{\nu m}^{(5)} = \lim_{\Omega\to0}\frac{i}{\Omega} T \sum_{l=-\infty}^{\infty} \int\frac{d^3\mathbf{k}}{(2\pi)^3}
\int \int d\omega d\omega^{\prime} \frac{\mbox{tr}\left[j_{\nu}(\mathbf{k})A(\omega; \mathbf{k})j_{m}^{5}(\mathbf{k})
A(\omega^{\prime}; \mathbf{k})\right]}{\left(i\omega_l+\mu-\omega\right)\left(i\omega_l-\Omega-i0+\mu-\omega^{\prime}\right)}.
\label{Kubo-E5-conductivity-calc-1}
\end{equation}
Then performing the summation over the Matsubara frequencies and setting $T=0$ at the end,
we derive the following result for the dissipative
\begin{eqnarray}
\label{Kubo-E5-conductivity-calc-rho}
\sigma_{0m}^{(5)} &=& -e^2\pi \int\frac{d^3\mathbf{k}}{(2\pi)^3} \frac{\delta_{\Gamma}^2(\mu-|\mathbf{d}|) -\delta_{\Gamma}^2(\mu+|\mathbf{d}|)}{|\mathbf{d}|} d_m\tilde{j}_m^{5},\\
\sigma_{nm}^{(5)} &=&-e^2\pi \int\frac{d^3\mathbf{k}}{(2\pi)^3}
\frac{1}{2|\mathbf{d}|^2}\sum_{s,s^{\prime}=\pm}
\delta_{\Gamma}(\mu-s|\mathbf{d}|)\delta_{\Gamma}(\mu-s^{\prime}|\mathbf{d}|)
\Bigg\{|\mathbf{d}|^2 \left(\partial_{k_n}d_m\right)\tilde{j}_m^5 \nonumber\\
&+&ss^{\prime}\sum_{i_1, i_2, i_3=1}^3\left[\delta_{i_1 i_2}\delta_{m i_3} -\delta_{i_1m}\delta_{i_2i_3}
+\delta_{i_1i_4}\delta_{i_2m}\right] (\partial_{k_n}d_{i_1})d_{i_2}\tilde{j}_m^5 d_{i_3}
\Bigg\},
\label{Kubo-E5-conductivity-calc-Re}
\end{eqnarray}
as well as nondissipative parts of the conductivity tensor
\begin{equation}
\tilde{\sigma}_{nm}^{(5)} = e\int\frac{d^3\mathbf{k}}{(2\pi)^3}
 \frac{\tilde{j}^{5}_m\left[\mathbf{d}\times(\partial_{k_n}\mathbf{d})\right]_m}{2|\mathbf{d}|^3} \left[1-\theta(|\mu|-|\mathbf{d}|)
 \right].
\label{Kubo-E5-conductivity-calc-Im-Gamma0}
\end{equation}
Here $\tilde{j}^5_n=\sum_{m=1}^3\mbox{tr}\left(\sigma_nj^5_m\right)/2$. Note that in the last equation we explicitly set $\Gamma\to0$. Formally, it is similar to the topological off-diagonal components of the conductivity tensor in Eq.~(\ref{Kubo-conductivity-calc-Im-Gamma0}).
Numerically, however, all spatial components vanish after the integration
over the whole Brillouin zone.

\begin{figure}[!ht]
\begin{center}
\includegraphics[width=0.5\textwidth]{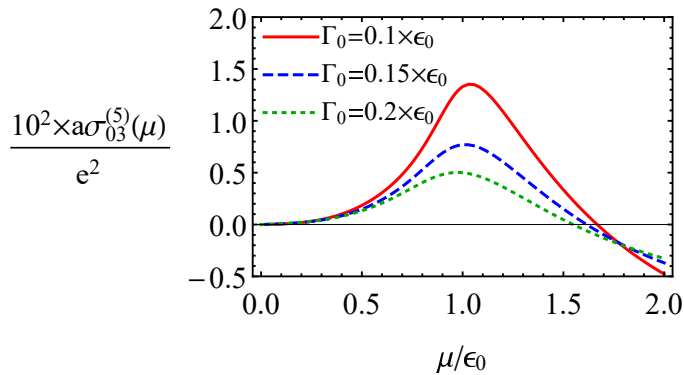}
\caption{The dependence of the conductivity tensor component $\sigma_{03}^{(5)}$ on
the chemical potential for three difference values of the quasiparticle decay width:
$\Gamma_0=0.1\,\epsilon_0$ (red solid line),
$\Gamma_0=0.15\,\epsilon_0$ (blue dashed line), and
$\Gamma_0=0.2\,\epsilon_0$ (green dotted line), respectively.
The numerical values of other parameters are defined in Appendix~\ref{Sec:App-model}.}
\label{fig:lattice-Kubo-E5}
\end{center}
\end{figure}

It is instructive, therefore, to investigate the conductivity tensor in
Eqs.~(\ref{Kubo-E5-conductivity-calc-rho}) and (\ref{Kubo-E5-conductivity-calc-Re}) in the case of a
nonzero quasiparticle width $\Gamma$. The corresponding calculations can be done
straightforwardly using numerical methods for a representative set of model parameters in
Appendix~\ref{Sec:App-model}. The analysis shows that $\sigma_{03}^{(5)}$ is the only nonzero
component of the conductivity tensor. The value of the corresponding component does not appear
to be protected by topology. This is clear from its dependence on the chemical potential shown
in Fig.~\ref{fig:lattice-Kubo-E5} for several choices of the quasiparticle width. Note that, by assumption, the
transport quasiparticle width includes a constant part $\Gamma_0$ as well as a frequency-dependent
part proportional to $\mu^2$ \cite{Burkov:2011}, i.e., $\Gamma(\mu)=\Gamma_0(1+\mu^2/\epsilon_0^2)$.
We would like to mention also that a nonzero $\sigma_{03}^{(5)}$ is quite interesting from a physics viewpoint. It represents
a form of dynamical piezoelectric effect that is driven by time-dependent strains in Weyl metals.

%%%%%%%%%%%%%%%%%%
%%%%%%%%%%%%%%%%%%

\section{Summary}
\label{sec:Summary-Discussions}

By making use of a generic lattice model of a Weyl semimetal, we calculated the electric charge and
current densities in the first order in background electromagnetic and strain-induced pseudoelectromagnetic
fields. A special attention in the analysis was paid to identifying the topological contributions
associated with the chiral properties of low-energy quasiparticles. In this connection, it should be
mentioned that the key features of the model (including the chiral anomaly) are captured
by its topology in the reciprocal space. Unlike the relativistic models in high-energy physics, which
are commonly used as a source of intuition for Weyl semimetals, lattice models require no
ultraviolet (high-energy) regularization and encounter no ambiguities in predicting physical observables.

Our results for the linear response in background electromagnetic and strain-induced pseudoelectromagnetic
fields show that, in addition to the usual matter part, there are two types of topological contributions in the
electric charge and current densities. The contributions of the first type are sensitive only to the Berry curvature
at Weyl nodes and can be reproduced exactly in the framework of the chiral kinetic theory. The contributions
of the second type are determined by a topological invariant (winding number), which is a global property
of the whole Brillouin zone, and cannot be captured by the chiral kinetic theory in its standard formulation.
Our direct calculations in the lattice model show that the contributions of the second type are given exactly
by the Bardeen-Zumino (or, equivalently, Chern-Simons) current. This finding reconfirms, therefore,
our claim in Ref.~\cite{Gorbar:2016ygi} that the physical definition of the electric current in the
consistent chiral kinetic theory must be amended by adding the Bardeen-Zumino term.

Our calculations indicate that the linear response in Weyl semimetals in the background pseudoelectromagnetic
fields, unlike its counterpart in the electromagnetic fields, is not expected to be completely universal.
Indeed, we found that even the formally topological part of the electric current induced by a constant
pseudomagnetic field in Eq.~(\ref{lattice-B5-J-n-mu}) has a nontrivial dependence on the model
parameters when the chemical potential is not very small. Such a discrepancy between the naive
expectation and the actual calculation can be explained by the fact that the strain-induced fields in
Weyl semimetals can be interpreted as the conventional pseudoelectromagnetic fields (i.e., introduced
via an axial vector potential $\mathbf{A}_5$ in the covariant derivatives) only in a close vicinity of the
Weyl nodes.

In this paper, we also showed that the same two types of topological currents are also induced in the case
of the multi-Weyl semimetals. Our direct calculations reveal, in fact, that the corresponding contributions
in the multi-Weyl semimetals contain an additional multiplication factor, which is the integer topological
charge of the Weyl nodes. This conclusion is not surprising and, in fact, agrees with a recent independent
analysis in Ref.~\cite{1705.04576}, where the high-energy inspired Fujikawa's regularization method was used.

Last but not least, by taking into account the topological origin of the winding number that determines the
Bardeen-Zumino terms, our calculation of the current in the lattice model also provides an instructive way of
justifying the definition of the consistent electric current in relativistic field theories. Indeed, the formulation
of the lattice model itself can be viewed as a form of regularizing a relativistic model that ensures the exact local
conservation of the electric charge. In this connection, one should remember, however, that the realization
of the chiral symmetry is nontrivial on a lattice \cite{Ginsparg-Kaplan}. Nevertheless, its implementation as
an anomalous symmetry in the low-energy effective theory might be sufficient for most practical purposes.

\begin{acknowledgments}
The work of E.V.G. was partially supported by the Program of Fundamental Research of the
Physics and Astronomy Division of the National Academy of Sciences of Ukraine.
The work of V.A.M. and P.O.S. was supported by the Natural Sciences and Engineering Research Council of Canada.
The work of I.A.S. was supported by the U.S. National Science Foundation under Grant No.~PHY-1404232.
\end{acknowledgments}

\appendix

\section{Model parameters}
\label{Sec:App-model}

In this appendix, we present a representative set of model parameters that we use in our
numerical calculations throughout the paper. In order to have a realistic model, we relate
the parameters in model (\ref{lattice-d-def-be}) to those in $\mathrm{Na_3Bi}$ using the
parametrization of Ref.~\cite{Wang}. The corresponding relations between the two sets
of model parameters read
\begin{eqnarray}
\label{lattice-coeff-C-be}
&&t_0=M_0-t_1-2t_2, \qquad t_{1,2}=-\frac{2M_{1,2}}{a^2},\\
&&g_0=C_0-g_1-2g_2, \qquad g_{1,2}=-\frac{2C_{1,2}}{a^2},\\
&&\Lambda=\frac{A}{a},
\label{lattice-coeff-C-ee}
\end{eqnarray}
where the numerical values of the new parameters are fixed by the band structure in $\mathrm{Na_3Bi}$
\cite{Wang},
\begin{equation}
\begin{array}{lll}
 C_0 = -0.06382~\mbox{eV},\qquad
& C_1 = 8.7536~\mbox{eV\,\AA}^2,\qquad
& C_2 = -8.4008~\mbox{eV\,\AA}^2,\\
 M_0=0.08686~\mbox{eV},\quad
& M_1=-10.6424~\mbox{eV\,\AA}^2,\qquad
& M_2=-10.3610~\mbox{eV\,\AA}^2,\\
 A=2.4598~\mbox{eV\,\AA}.
\end{array}
\label{lattice-model-parameters}
\end{equation}
For the sake of simplicity, in this paper, we assume that the Weyl semimetal model has a cubic
lattice, i.e., $a_x=a_y=a_z=a=7.5~\mbox{\AA}$. Although typically this is not the case in real
materials, there are no important topological consequences resulting from such an assumption.

\section{Multi-Weyl semimetals}
\label{Sec:App-multi-Weyl}

In this appendix, we discuss how to define a lattice Hamiltonian for multi-Weyl materials by using
the same general model as in Eq.~(\ref{lattice-d-def-be}).

By definition, the multi-Weyl semimetals are Weyl semimetals with the topological charges of
Weyl nodes greater than one. The low-energy effective Hamiltonian for the multi-Weyl semimetal
can be given in the following form \cite{1705.04576,Volovik:1988,Fang-Bernevig:2012,Li-Roy-Das-Sarma:2016}:
\begin{equation}
\mathcal{H}_{\lambda} = \lambda v_F \left[(\mathbf{p}+\lambda\mathbf{b})_3 \sigma_z + (\mathbf{p}+\lambda\mathbf{b})_{+}^{n}\sigma_{-} + (\mathbf{p}+\lambda\mathbf{b})_{-}^{n}\sigma_{+}\right],
\label{lattice-H-multi-lin-def}
\end{equation}
where $n=2,3$ is the topological charge of the Weyl nodes, $\lambda=\pm$ is the chirality,
$\mathbf{p}_{\pm}=\left(p_{x}\pm ip_{y}\right)/\sqrt{2}$,
$\sigma_{\pm}=\left(\sigma_{x}\pm i\sigma_{y}\right)/\sqrt{2}$, and
$\mathbf{b}$ is the chiral shift parameter that defines the momentum space separation between
the Weyl nodes. It is straightforward to check that the corresponding lattice formulation of the
Hamiltonian can be given by the same Eq.~(\ref{lattice-d-def-be}), but with a different choice of
functions $d_1$ and $d_2$. In the case of Weyl nodes with the topological charge $n=2$, for
example, one can use the following choice of functions:
\begin{eqnarray}
\label{d-def-n=2-be}
d_1 &=& \Lambda \frac{\sin^2{(a_xk_x)}-\sin^2{(a_yk_y)}}{\sqrt{2}},\\
d_2 &=& \Lambda \frac{\sin{(a_xk_x)}\sin{(a_yk_y)}}{\sqrt{2}}.
\label{d-def-n=2-ee}
\end{eqnarray}
Similarly, in the case of the Weyl nodes with the topological charge $n=3$, one
can use
\begin{eqnarray}
\label{d-def-n=3-be}
d_1 &=& \Lambda \frac{\sin^3{(a_xk_x)}-3\sin{(a_xk_x)}\sin^2{(a_yk_y)}}{2},\\
d_2 &=& -\Lambda \frac{\sin^3{(a_yk_y)}-3\sin{(a_yk_y)}\sin^2{(a_xk_x)}}{2}.
\label{d-def-n=3-ee}
\end{eqnarray}

\section{Integrals over $\omega$}
\label{Sec:App-Int-omega}

In this appendix, we present the results for several types of integrals over $\omega$ that we encounter
in the calculation of the linear response in the main text, as well as in other appendixes. By omitting the
intermediate steps of derivations, here we give only the final results:
\begin{eqnarray}
\label{lattice-zero-mu-eps0-omega-int-be}
I_1&=&\int \frac{d\omega}{ \left[\omega-d_0+i0\sign{\omega}\right]^2-|\mathbf{d}|^2 }
=  -\frac{\pi i}{|\mathbf{d}|}\left[1 - \theta(|d_0|-|\mathbf{d}|)\right],\\
\label{lattice-zero-mu-eps0-omega-int-ee}
I_2&=&\int \frac{(\omega-d_0)\, d\omega}{ \left[\omega-d_0+i0\sign{\omega}\right]^2-|\mathbf{d}|^2 }
=  -\pi i \,\sign{d_0}\theta(|d_0|-|\mathbf{d}|), \\
\label{mu-omega-int-be}
I_3&=&\int \frac{d\omega}{\left\{\left[\omega-d_0+i0\sign{\omega}\right]^2-|\mathbf{d}|^2\right\}^2} = \frac{\pi i}{|\mathbf{d}|}\,\left\{\frac{1}{2|\mathbf{d}|^2}
\left[1-\theta\left(|d_0|-|\mathbf{d}|\right)\right] -\delta\left(d_0^2-|\mathbf{d}|^2\right)\right\}, \\
I_4&=&\int \frac{(\omega-d_0)\, d\omega}{\left\{\left[\omega-d_0+i0\sign{\omega}\right]^2-|\mathbf{d}|^2\right\}^2} = \pi i\,
\sign{d_0}\left[\delta\left(d_0^2-|\mathbf{d}|^2\right) -\delta\left(-|\mathbf{d}|^2\right)\right] , \\
I_5&=&\int \frac{(\omega-d_0)^2 d\omega}{\left\{\left[\omega-d_0+i0\sign{\omega}\right]^2-|\mathbf{d}|^2\right\}^2} = -\pi i\,\left\{\frac{1}{2|\mathbf{d}|}
\left[1-\theta\left(|d_0|-|\mathbf{d}|\right)\right] +|d_0|\delta\left(d_0^2-|\mathbf{d}|^2\right) \right\},
\label{mu-omega-int-ee}
\end{eqnarray}
where $\theta(x)$ is the unit step function. Note that the above integrals are straightforward to calculate by
using the following relations obtained in Ref.~\cite{Gorbar:2013upa}:
\begin{eqnarray}
\frac{1}{\left\{\left[\omega-d_0+i0\sign{\omega}\right]^2-|\mathbf{d}|^2\right\}^n} &=& \frac{1}{\left[\left(\omega-d_0\right)^2-|\mathbf{d}|^2+i0\right]^n}
\nonumber\\
&+&\frac{2\pi i (-1)^{n-1}}{(n-1)!}\theta\left(|d_0|-|\omega|\right)\theta\left(\omega d_0\right) \delta^{(n-1)}\left[(\omega-d_0)^2-|\mathbf{d}|^2\right],
\\
\frac{2n\left(\omega-d_0\right)}{\left\{\left[\omega-d_0+i0\sign{\omega}\right]^2-|\mathbf{d}|^2\right\}^{n+1}} &=&-\frac{\partial}{\partial \omega}
\left(\frac{1}{\left\{\left[\omega-d_0+i0\sign{\omega}\right]^2-|\mathbf{d}|^2\right\}^{n}}\right) \nonumber\\
&-&\frac{2\pi i (-1)^{n}}{(n-1)!}\sign{d_0} \delta^{(n-1)}\left[(\omega-d_0)^2-|\mathbf{d}|^2\right]\left[\delta(\omega)-\delta(\omega-d_0)\right].
\label{Lin-omega-ints-formulas}
\end{eqnarray}

\section{The Green's function in the first order in magnetic field}
\label{Sec:App-Green-first}

In this appendix, we present the details of the Green's function calculation in the first order
in a constant background magnetic field. Because of the dependence of the vector potential on the spatial
coordinate(s), e.g., in the Landau gauge $A_{\nu}(\mathbf{r})=\left(0,0,-xB,0\right)$, the translation
invariance is formally broken and the calculation of the first-order correction to the Green's function
$G^{(1)}(r,r^{\prime})$ becomes rather nontrivial. Indeed, while it is natural to use the momentum-space
representation for the translation invariant zeroth-order Green's function, the spatial dependence in
the gauge field $A_{\nu}(\mathbf{r})$ (which enters through the interaction Hamiltonian) complicates
the analysis.

In order to partially circumvent the technical complications associated with the absence of translation
invariance, it is convenient to utilize the following representation for the spatial coordinate:
\begin{equation}
x^{\prime \prime} = \int dp_x \left[-i\partial_{p_x}e^{ip_x x^{\prime \prime}}\right] \delta(p_x)
= \int dp_x e^{ip_x x^{\prime \prime}} i[\partial_{p_x}\delta(p_x)].
\end{equation}
Then, by using Eq.~(\ref{first-order-correction}), we can rewrite the first-order correction to the Green's
function in the form:
\begin{eqnarray}
G^{(1)}(r,r^{\prime}) &=& -\frac{B}{2} \int \frac{d\omega d^3\mathbf{k}}{(2\pi)^4}  e^{-i\omega(t-t^{\prime}) +i\mathbf{k}(\mathbf{r} -\mathbf{r}^{\prime})} \Big\{i(x-
x^{\prime})G^{(0)}(\omega; \mathbf{k})j_{y}(\mathbf{k}) G^{(0)}(\omega,\mathbf{k}) +2ix^{\prime}G^{(0)}(\omega; \mathbf{k})j_{y}(\mathbf{k}) G^{(0)}(\omega;\mathbf{k})\nonumber\\
&+& \left[\partial_{k_x}G^{(0)}(\omega; \mathbf{k})\right] j_{y}(\mathbf{k}) G^{(0)}(\omega;\mathbf{k}) - G^{(0)}(\omega; \mathbf{k})j_{y}
(\mathbf{k}) \left[\partial_{k_x}G^{(0)}(\omega; \mathbf{k})\right]\Big\}.
\label{lattice-first-G1-01}
\end{eqnarray}
Because of the dependence on the relative coordinate $x-x^{\prime}$ in the first term in the square brackets, a special care should
be taken when using this in the calculation of the current density defined by Eq.~(\ref{lattice-J-def}).
Indeed, the definition also contains the current operator $\mathbf{j}(-i\partial_{\mathbf{r}})$
that acts not only on the phase factor $e^{i\mathbf{k}(\mathbf{r} -\mathbf{r}^{\prime})}$, but also
on $(r_n-r_n^{\prime})$. The corresponding result can be calculated systematically by using a
series representation for the trigonometric functions in the definition of $\mathbf{j}(-i\partial_{\mathbf{r}})$;
see Eqs.~(\ref{lattice-coeff-j-be})--(\ref{lattice-coeff-j-ee}) with $\mathbf{k}$ replaced
by $-i\partial_{\mathbf{r}}$. Thus, by making use of the relations
\begin{eqnarray}
\sin{(-ia_n\partial_n)} e^{ik_nr_n}r_n \Big|_{r_n\to0}&=& \sum_{m=0}^{\infty} (-1)^{n} (-ia_n)^{2m+1} \frac{(ik_n)^{2m}}{2m!} = (-ia_n)
\cos{(a_nk_n)},\\
\cos{(-ia_n\partial_n)} e^{ik_nr_n}r_n \Big|_{r_n\to0}&=& \sum_{m=1}^{\infty} (-1)^{n} (-ia_n)^{2m} \frac{(ik_n)^{2m-1}}{(2m-1)!} = -(-ia_n)
\sin{(a_nk_n)},
\end{eqnarray}
we derive the following result for the current density due to the first term in Eq.~(\ref{lattice-first-G1-01}):
\begin{eqnarray}
&&-\frac{B}{2}\lim_{r^{\prime}\to r}\int \frac{d\omega d^3\mathbf{k}}{(2\pi)^4}j_{n}(-i\partial_{\mathbf{r}})i(r_n-r_n^{\prime})G^{(0)}(\omega; \mathbf{k})j_{y}(\mathbf{k}) G^{(0)}
(\omega;\mathbf{k}) e^{-i\omega(t-t^{\prime}) +i\mathbf{k}(\mathbf{r} -\mathbf{r}^{\prime})} \nonumber\\
&&= -\frac{B}{2}\int \frac{d\omega d^3\mathbf{k}}{(2\pi)^4}(\partial_{k_n}j_n(\mathbf{k}))G^{(0)}(\omega; \mathbf{k})j_{y}(\mathbf{k}) G^{(0)}(\omega;\mathbf{k}).
\label{lattice-first-current-repl}
\end{eqnarray}
By combining this with the two additional contributions due to the other three terms in
the first-order Green's function in Eq.~(\ref{lattice-first-G1-01}), we obtain the final
result for the current density presented in Eq.~(\ref{lattice-first-J}) in the main text.

\section{Derivation of the electric charge and current densities in background electromagnetic fields}
\label{sec:App-key-point-charge-current}

In this appendix, we provide the detailed derivations of the electric charge and current densities
in background magnetic and electric fields.

\subsection{Background magnetic field}
\label{sec:App-key-point-charge-current-B}

Let us start with the case of a background magnetic field. The electric charge density in Eq.~(\ref{lattice-first-rho}) can be
rewritten as follows:
\begin{equation}
\rho = \frac{eB}{2} \int \frac{d\omega d^3\mathbf{k}}{(2\pi)^4} \frac{1}{N^2}\mbox{tr}\Bigg[(\partial_{k_x}Q) j_{y}(\mathbf{k})Q -
\frac{1}{N}(\partial_{k_x}N) Qj_{y}(\mathbf{k})Q - Qj_{y}(\mathbf{k}) (\partial_{k_x}Q) +\frac{1}{N}Qj_{y}(\mathbf{k})Q (\partial_{k_x}N)
\Big],
\end{equation}
where we used the following shorthand notations for the numerator and denominator of the zeroth-order
Green's function $G^{(0)}(\omega; \mathbf{k})$:
\begin{eqnarray}
Q&\equiv&i\left[\omega+\mu+(\mathbf{d}\cdot\bm{\sigma})\right],\\
N&\equiv&\left[\omega+\mu+i0\sign{\omega}\right]^2-\mathbf{d}^2.
\end{eqnarray}
After some algebraic simplifications, this can be rewritten in the following simple form:
\begin{equation}
\rho =\frac{eB}{2} \int \frac{d\omega d^3\mathbf{k}}{(2\pi)^4} \sum_{i_1, i_2, i_3=1}^{3}\frac{4i\epsilon_{i_1i_2i_3}}{N^2} \left(\partial_{k_x}d_{i_1}\right)\left(\partial_{k_y}d_{i_2}\right)d_{i_3}.
\label{lattice-topology-rho}
\end{equation}
Finally, performing the integration over $\omega$ using the result in Eq.~(\ref{mu-omega-int-be}),
we obtain the topological and matter parts of the charge density in Eqs.~(\ref{lattice-topology-rho-1})
and (\ref{lattice-topology-mu-rho-01}), respectively.

In the case of the electric current density $J_n$, where $n=1,2,3$, we use the definition in
Eq.~(\ref{lattice-first-J}). In the first order in an external magnetic field, it gives
\begin{eqnarray}
J_n&=&-\frac{e^2B}{2}\int \frac{d\omega d^3\mathbf{k}}{(2\pi)^4} \frac{1}{N^2}\mbox{tr}\Bigg\{(\partial_{k_n}Q)(\partial_{k_x}Q)
(\partial_{k_y}Q)Q  - (\partial_{k_n}Q)Q(\partial_{k_y}Q)(\partial_{k_x}Q)  \nonumber\\
&+&\delta_{n,x}(\partial_{k_x}^2Q)Q(\partial_{k_y}Q)Q
+2irx\delta_{n,x}(\partial_{k_x}Q)Q(\partial_{k_y}Q)Q \Bigg\}.
\label{lattice-topology-J-n}
\end{eqnarray}
The first two terms in the curly brackets can be combined to give
\begin{eqnarray}
J_n^{(1)}= -\frac{e^2B}{2}\int \frac{d\omega d^3\mathbf{k}}{(2\pi)^4} \frac{\omega+\mu}{N^2} \sum_{i_1,i_2,i_3=1}^3
4i\epsilon_{i_1i_2i_3} (\partial_{k_n}d_{i_1})(\partial_{k_x}d_{i_2})(\partial_{k_y}d_{i_3}).
\label{lattice-topology-J-n-term-1}
\end{eqnarray}
The last two terms in the curly brackets in Eq.~(\ref{lattice-topology-J-n}) can be rewritten as follows:
\begin{eqnarray}
J_n^{(2)}&=& -\delta_{n,x} e^2B \int \frac{d\omega d^3\mathbf{k}}{(2\pi)^4} \frac{(\omega+\mu)^2}{N^2}
\left[\left((\partial_{k_x}^2\mathbf{d})\cdot(\partial_{k_y}\mathbf{d})\right) +2ix\left((\partial_{k_x}\mathbf{d})\cdot(\partial_{k_y}\mathbf{d})\right)\right]\nonumber\\
&-&\delta_{n,x}e^2B \int \frac{d\omega d^3\mathbf{k}}{(2\pi)^4} \frac{1}{N^2} \sum_{i_1, i_2, i_3, i_4=1}^3\left[\delta_{i_1i_2}\delta_{i_3i_4}-\delta_{i_1i_3}\delta_{i_2i_4}
+\delta_{i_1i_4}\delta_{i_2i_3}\right] \left[(\partial_{k_x}^2d_{i_1}) +2ix(\partial_{k_x}d_{i_1})\right]d_{i_2}(\partial_{k_y}d_{i_3})d_{i_4}.\nonumber\\
\label{lattice-topology-J-n-term-23}
\end{eqnarray}
Then, we integrate over $\omega$ by using Eqs.~(\ref{mu-omega-int-be})--(\ref{mu-omega-int-ee}) and
arrive at the following result:
\begin{eqnarray}
J_n&=&  -e^2B \int \frac{d^3\mathbf{k}}{(2\pi)^3} \sign{\mu} \sum_{i_1,i_2,i_3=1}^{3}\epsilon_{i_1i_2i_3}
(\partial_{k_n}d_{i_1})(\partial_{k_x}d_{i_2})(\partial_{k_y}d_{i_3}) \delta\left(\mu^2-|\mathbf{d}|^2\right) \nonumber\\
&+&i\delta_{n,x} e^2B \int \frac{d^3\mathbf{k}}{2(2\pi)^3} \left[\left((\partial_{k_x}^2\mathbf{d})\cdot(\partial_{k_y}\mathbf{d})\right) +2ix\left((\partial_{k_x}\mathbf{d})\cdot(\partial_{k_y}\mathbf{d})\right)\right]
\left\{\frac{1}{2|\mathbf{d}|}\left[1-\theta\left(|\mu|-|\mathbf{d}|\right)\right] +|\mu|\delta\left(\mu^2-|\mathbf{d}|^2\right) \right\} \nonumber\\
&-&i\delta_{n,x} e^2B \int \frac{d^3\mathbf{k}}{2(2\pi)^3} \frac{1}{|\mathbf{d}|} \sum_{i_1, i_2, i_3, i_4=1}^3\left[\delta_{i_1i_2}\delta_{i_3i_4}-\delta_{i_1i_3}\delta_{i_2i_4}
+\delta_{i_1i_4}\delta_{i_2i_3}\right] \left[(\partial_{k_x}^2d_{i_1}) +2ix(\partial_{k_x}d_{i_1})\right]d_{i_2}(\partial_{k_y}d_{i_3})d_{i_4} \nonumber\\
&\times&\left\{\frac{1}{2|\mathbf{d}|^2}
\left[1-\theta\left(|\mu|-|\mathbf{d}|\right)\right] -\delta\left(\mu^2-|\mathbf{d}|^2\right)\right\}.%\nonumber\\
\label{lattice-topology-J-n-1-app}
\end{eqnarray}
Note that the last two parts vanish after the integration over the whole
Brillouin zone.

\subsection{Background electric field}
\label{sec:App-key-point-charge-current-E}

In this subsection we present the details of derivation of the conductivity tensor (\ref{Kubo-conductivity-calc-1})
in Sec.~\ref{sec:Kubo}. By performing the summation over the Matsubara frequencies in the corresponding
expression, we arrive at the conventional representation for the DC  conductivity tensor in terms of the
spectral function:
\begin{equation}
\sigma_{\nu m} = \mbox{Re}{\left(\lim_{\Omega\to0}\frac{i}{\Omega} \int\frac{d^3\mathbf{k}}{(2\pi)^3}
\int \int d\omega d\omega^{\prime} \frac{n_F(\omega)-n_F(\omega^{\prime})}{\omega-\omega^{\prime}-\Omega-i0}\mbox{tr}
\left[j_{\nu}(\mathbf{k})A(\omega; \mathbf{k})\mathbf{j}_{m}(\mathbf{k})A(\omega^{\prime}; \mathbf{k})\right]\right)},
\label{Kubo-conductivity-calc-2}
\end{equation}
where $n_{F}(\omega)=1/\left[e^{(\omega-\mu)/T}+1\right]$ is the Fermi-Dirac distribution, $\nu=0,1,2,3$,
and the spectral function $A(\omega; \mathbf{k})$ is defined in Eq.~(\ref{Kubo-spectral-function-def}).

In the case of $\nu=0$, the trace in the expression for the conductivity reads
\begin{equation}
\mbox{tr}\left[j_{0}(\mathbf{k})A(\omega; \mathbf{k})j_{m}(\mathbf{k})A(\omega^{\prime}; \mathbf{k})\right] =\frac{e^2}{2|\mathbf{d}|^2}
\sum_{s,s^{\prime}=\pm}ss^{\prime}\delta_{\Gamma}(\omega-s|\mathbf{d}|)\delta_{\Gamma}(\omega^{\prime}-s^{\prime}|\mathbf{d}|)
|\mathbf{d}|(s+s^{\prime}) \left(\mathbf{d}\cdot\partial_{k_m}\mathbf{d}\right),
\label{Kubo-nu=0-trace}
\end{equation}
where $\delta_{\Gamma}(x)$ is a regularized form of the $\delta$-function defined in
Eq.~(\ref{Kubo-d-Gamma}). By substituting this in Eq.~(\ref{Kubo-conductivity-calc-2}),
extracting the real part with the help of the Sokhotski formula, and integrating over
$\omega^{\prime}$, we obtain
\begin{eqnarray}
\sigma_{0m} &=& e^2\pi \int\frac{d^3\mathbf{k}}{(2\pi)^3}
\int d\omega \frac{1}{4T\cosh^2{\left(\frac{\omega-\mu}{2T}\right)}} \frac{1}{2|\mathbf{d}|} \sum_{s,s^{\prime}=\pm}ss^{\prime}\delta_{\Gamma}(\omega-s|\mathbf{d}|)\delta_{\Gamma}(\omega-s^{\prime}|\mathbf{d}|)(s+s^{\prime})
\left(\mathbf{d}\cdot(\partial_{k_m}\mathbf{d})\right) \nonumber\\
&\stackrel{T\to0}{=}&e^2\pi \int\frac{d^3\mathbf{k}}{(2\pi)^3} \frac{\delta_{\Gamma}^2(\mu-|\mathbf{d}|) -\delta_{\Gamma}^2(\mu+|\mathbf{d}|)}{|\mathbf{d}|}
\left(\mathbf{d}\cdot(\partial_{k_m}\mathbf{d})\right).
\label{Kubo-conductivity-calc-Im-app}
\end{eqnarray}

In the case of $\nu=n$, the result for the trace in Eq.~(\ref{Kubo-conductivity-calc-2}) is given by
the following expression:
\begin{equation}
\mbox{tr}\left[j_{n}(\mathbf{k})A(\omega; \mathbf{k})j_{m}(\mathbf{k})A(\omega^{\prime}; \mathbf{k})\right]
= -\frac{e^2}{4|\mathbf{d}|^2}\sum_{s,s^{\prime}=\pm}ss^{\prime} \delta_{\Gamma}(\omega-s|\mathbf{d}|)\delta_{\Gamma}(\omega^{\prime}-s^{\prime}|\mathbf{d}|)
\left[i|\mathbf{d}|(s^{\prime}-s)T_{1}+T_{2}(s, s^{\prime})\right],
\label{Kubo-trace}
\end{equation}
where we introduced the shorthand notations,
\begin{equation}
T_{1} \equiv 2\sum_{i_1, i_2, i_3=1}^3\epsilon_{i_1 i_2 i_3}(\partial_{k_n}d_{i_1})d_{i_2}(\partial_{k_m}d_{i_3}),
\end{equation}
\begin{equation}
T_{2}(s, s^{\prime}) \equiv 2ss^{\prime}|\mathbf{d}|^2 \left((\partial_{k_n}\mathbf{d})\cdot(\partial_{k_m}\mathbf{d})\right) +2\sum_{i_1, i_2, i_3, i_4=1}^3\left[\delta_{i_1 i_2}\delta_{i_3i_4} -\delta_{i_1i_3}\delta_{i_2i_4}
+\delta_{i_1i_4}\delta_{i_2i_3}\right] (\partial_{k_n}d_{i_1})d_{i_2}(\partial_{k_m}d_{i_3})d_{i_4}.
\label{Kubo-trace-1}
\end{equation}
It is convenient to calculate separately the contributions to the conductivity tensor originating from
the two different terms in the square brackets in Eq.~(\ref{Kubo-trace}). By using the definition in
Eq.~(\ref{Kubo-conductivity-calc-2}), we rewrite the contribution due to the first term as follows:
\begin{equation}
\sigma_{nm}^{(1)} = -\mbox{Re}{\left(e^2\lim_{\Omega\to0}\frac{i}{\Omega} \int\frac{d^3\mathbf{k}}{(2\pi)^3}
\int \int d\omega d\omega^{\prime} \frac{n_F(\omega)-n_F(\omega^{\prime})}{\omega-\omega^{\prime}-\Omega}
\frac{1}{2|\mathbf{d}|}\Big[\delta_{\Gamma}(\omega+|\mathbf{d}|)\delta_{\Gamma}(\omega^{\prime}-|\mathbf{d}|) -\delta_{\Gamma}(\omega-|\mathbf{d}|)\delta_{\Gamma}(\omega^{\prime}+|\mathbf{d}|)\Big]
iT_{1} \right)}.
\label{Kubo-conductivity-calc-Im}
\end{equation}
In order to extract the topological part of the conductivity, we consider the clean limit in
Eq.~(\ref{Kubo-conductivity-calc-Im}). Thus, by setting $\Gamma\to0$
and integrating
over $\omega$, we arrive at
\begin{equation}
\sigma_{nm}^{(1)}= e^2\lim_{\Omega\to0}\frac{1}{\Omega} \int\frac{d^3\mathbf{k}}{(2\pi)^3}\frac{\Omega}{|\mathbf{d}|}  \left[n_F(-|\mathbf{d}|)-n_F(|\mathbf{d}|)\right] \frac{T_{1}}{4|\mathbf{d}|^2-\Omega^2}
=e^2 \int\frac{d^3\mathbf{k}}{(2\pi)^3}
\frac{T_{1}}{4|\mathbf{d}|^3} \left[n_F(-|\mathbf{d}|)-n_F(|\mathbf{d}|)\right].
\end{equation}
Taking the limit $T \to 0$, one can easily obtain Eq.~(\ref{Kubo-conductivity-calc-Im-Gamma0}).

Similarly, the contribution to the conductivity tensor due to the second term in the square brackets in
Eq.~(\ref{Kubo-trace}) is given by
\begin{eqnarray}
\sigma_{nm}^{(2)} &=& e^2\lim_{\Omega\to0}\frac{i}{\Omega} \int\frac{d^3\mathbf{k}}{(2\pi)^3}
\int \int d\omega d\omega^{\prime} \left[n_F(\omega)-n_F(\omega^{\prime})\right] (\pi\,i)\delta{\left(\omega-\omega^{\prime}-\Omega\right)} \frac{1}{4|\mathbf{d}|^2}\sum_{s,s^{\prime}=\pm} s s^{\prime}
\delta_{\Gamma}(\omega-s|\mathbf{d}|)\delta_{\Gamma}(\omega^{\prime}-s^{\prime}|\mathbf{d}|)
\nonumber\\
&\times&
T_{2}(s,s^{\prime})
=e^2\pi \int\frac{d^3\mathbf{k}}{(2\pi)^3}
\int d\omega \frac{1}{4T\cosh^2{\left(\frac{\omega-\mu}{2T}\right)}} \frac{1}{4|\mathbf{d}|^2}\sum_{s,s^{\prime}=\pm}s s^{\prime}
\delta_{\Gamma}(\omega-s|\mathbf{d}|)\delta_{\Gamma}(\omega-s^{\prime}|\mathbf{d}|)T_{2}(s,s^{\prime}).
\end{eqnarray}
After integrating over $\omega$ and taking the limit $T\to 0$, the above expression reduces to
Eq.~(\ref{Kubo-conductivity-calc-Re}).

\end{document}